\documentclass[%
superscriptaddress,
 amsmath,amssymb,
 aps,
twocolumn,
prb,
]{revtex4-2}

\usepackage{graphicx}
\usepackage{dcolumn}
\usepackage{bm}
\usepackage{physics}
\usepackage[colorlinks=true, citecolor=red, linkcolor=blue, urlcolor=black]{hyperref}
\usepackage{here}

\usepackage[T1]{fontenc}
\usepackage[utf8]{inputenc}

\begin{document}


\title{Existence of two distinct valence bond solid states in the dimerized frustrated ferromagnetic  $J_1$-$J_1'$-$J_2$ chain}

\author{J\k{e}drzej Wardyn}

 \affiliation{Faculty of Physics, University of Warsaw, Pasteura 5, PL-02093 Warsaw, Poland}
 \affiliation{Institute for Theoretical Solid State Physics, IFW Dresden, 01171 Dresden, Germany}
\author{Satoshi Nishimoto}
\affiliation{Institute for Theoretical Solid State Physics, IFW Dresden, 01171 Dresden, Germany}
\affiliation{Department of Physics, Technical University Dresden, 01069 Dresden, Germany}
\author{Cli\`o Efthimia Agrapidis}
\email{clio.agrapidis@fuw.edu.pl}
\affiliation{Faculty of Physics, University of Warsaw, Pasteura 5, PL-02093 Warsaw, Poland}

\date{\today}

\begin{abstract}
We study the frustrated and dimerized ferromagnetic-antiferromagnetic $J_1$-$J_1'$-$J_2$ chain using the density-matrix renormalization group method. Based on numerical calculations of the second derivative of energy, spin gap, spin-spin correlations, string order parameter (SOP), and entanglement spectrum (ES), we obtain the ground-state phase diagram for a wide range of $J_1'/J_1$ and $J_2/|J_1|$ values. This phase diagram reveals a ferromagnetic phase and two distinct valence-bond-solid (VBS) phases. The first VBS phase, referred to as $\mathcal{D}_3$-VBS, is typified by the formation of valence bonds between third-neighbor spin-1/2's, persisting as a continuation from the $J_1'/J_1=1$ limit. Alternatively, the second VBS phase, referred to as mixed-VBS, exhibits a coexistence of both second- and third-neighbor valence bonds, interpreted as a continuation from the $J_1'/J_1=0$ case. Remarkably, both VBS states are identified as being of Haldane-type, marked by a finite SOP and 2-fold ES degeneracy. Unexpectedly, our analysis uncovers a significant enhancement of the valence bond stability at the boundary of the two VBS phases. This study provides the first empirical demonstration of a nontrivial quantum phase transition between different topological VBS states in spin-1/2 chains. Moreover, we find that the ground state of the relevant quasi-one-dimensional material LiCuSbO$_4$ is classified as the $\mathcal{D}_3$-VBS state. Collectively, these results mark a substantial stride forward in our comprehension of quantum phase transitions and topological states.
\end{abstract}

\maketitle


\section{\label{sec:Introduction} Introduction}

The interplay of magnetic frustration and low-dimensionality can lead to exotic states of matter in spin chains due to extensive quantum fluctuations \cite{Balents2010,Moessner2006}. A simple representative of such systems is the one-dimensional (1D) $J_1$-$J_2$ model, where $J_1$ and $J_2$ represent the nearest-neighbor and next-nearest-neighbor magnetic interactions, respectively. This model has facilitated understanding the magnetic behavior of various materials, including certain quasi-1D magnetic insulators like copper-based compounds. While the signs of the interactions can be influenced by a plethora of factors -- including the crystal structure, electronic configuration as well as the nature of the magnetic ions and their ligands -- they can be broadly categorized as follows \cite{Kuzian2023}: In corner-shared cuprates, both $J_1$ and $J_2$ interactions are antiferromagnetic (AFM) because the superexchange interaction between the copper ions is mediated through an oxygen atom at the corner. Whereas in edge-shared cuprates, the nearest-neighbor $J_1$ interaction is typically ferromagnetic (FM), while the next-nearest-neighbor $J_2$ interaction remains AFM. The FM $J_1$ interaction comes about due to the direct overlap of the copper and oxygen orbitals along the edge. Therefore, the 1D $J_1$-$J_2$ model is straightforward enough to be studied theoretically and numerically, but it offers a good starting point to comprehend the intricate physics in real materials, especially regarding frustration and magnetic ordering.

The ground state of the AFM-AFM $J_1$-$J_2$ chain at $J_2 \gtrsim 0.241$ is identified as a valence bond solid (VBS) state, as defined by the Majumdar-Ghosh picture \cite{Majumdar1969}. In contrast, recent findings suggest that the ground state of the FM-AFM $J_1$-$J_2$ chain exhibits a distinct type of VBS state, characterized by the formation of a valence bond dimer between third-neighboring spin-1/2's, referred to as the $\mathcal D_3$-VBS state \cite{Agrapidis2019}. This state is especially intriguing as it represents a topologically non-trivial phase, bearing similarities with the Haldane \cite{Haldane1983a,Haldane1983b,Botet1983,Renard1988} or Affleck-Kennedy-Lieb-Tasaki (AKLT)-like VBS state \cite{Affleck1987,Affleck1988}. The emergence of this state can be intuitively anticipated due to the emergence of effective spin-1 degrees of freedom, resulting from the spontaneous formation of nearest-neighbor FM dimers of spin-1/2's.

The current work extends the rudimentary chain model to consider the dimerized frustrated FM $J_1$-$J_1'$-$J_2$ chain. This model introduces a further complication of different types of nearest-neighbor interactions, $J_1$ and $J_1'$ ($|J_1|>|J_1'|$), which alternate in a dimerized pattern. The presence of this structural FM alternation was indeed predicted in the edge-shared cuprate LiCuSbO$_4$ \cite{Grafe2017}. However, it is largely unknown how the ground state of this model changes as a function of dimerization and frustration. From a theoretical perspective, the role of such alternating interactions may even simplify the VBS issue, given that the effective spin-1 degrees of freedom are explicitly created on the $J_1$ bond. This is equivalent to the formation of a spin-triplet dimer on the $J_1$ bond. In the strong dimer limit $J_1'/J_1=0$, the $J_1$-$J_1'$-$J_2$ chain can be effectively mapped onto the spin-1 Heisenberg chain \cite{Hida1991,Watanabe1993} (see also Sec. \ref{sec:model}). This indicates that the ground state is no longer the $\mathcal D_3$-VBS one at $J_1'/J_1=0$. Thus, a crossover or phase transition is feasible between the $\mathcal D_3$-VBS state at $J_1'/J_1=1$ and a symmetry-protected Haldane state at $J_1'/J_1=0$. 

Prompted by this, we examine the $J_1$-$J_1'$-$J_2$ chain using the density-matrix renormalization group (DMRG) algorithm \cite{White1993, Schollwoeck2005}. We first detect the presence of phase boundary by observing the second derivative of the ground-state energy as a function of the parameters. We then calculate the spin gap and string order parameter (SOP) to confirm the presence of a VBS state. Earlier, some of the present authors had estimated the spin gap for this system \cite{Agrapidis2017}. However, in certain specific parameter regions, it was somewhat underestimated due to insufficient management of the edge state, since the VBS structure was not well established. Consequently, it is revisited in this work. Lastly, local spin-spin correlations and the entanglement spectra (ES) are considered to further validate the estimated phase boundary.

The remainder of the paper is organized as follows: Section II elaborates on the model and details the applied numerical method. Section III presents the ground-state phase diagram based on the numerical results. Section IV is devoted to summarizing the study and discussing the findings.

\section{Model and Methods}\label{sec:model}

\begin{figure}[t]
	\centering
	\includegraphics[width=\columnwidth]{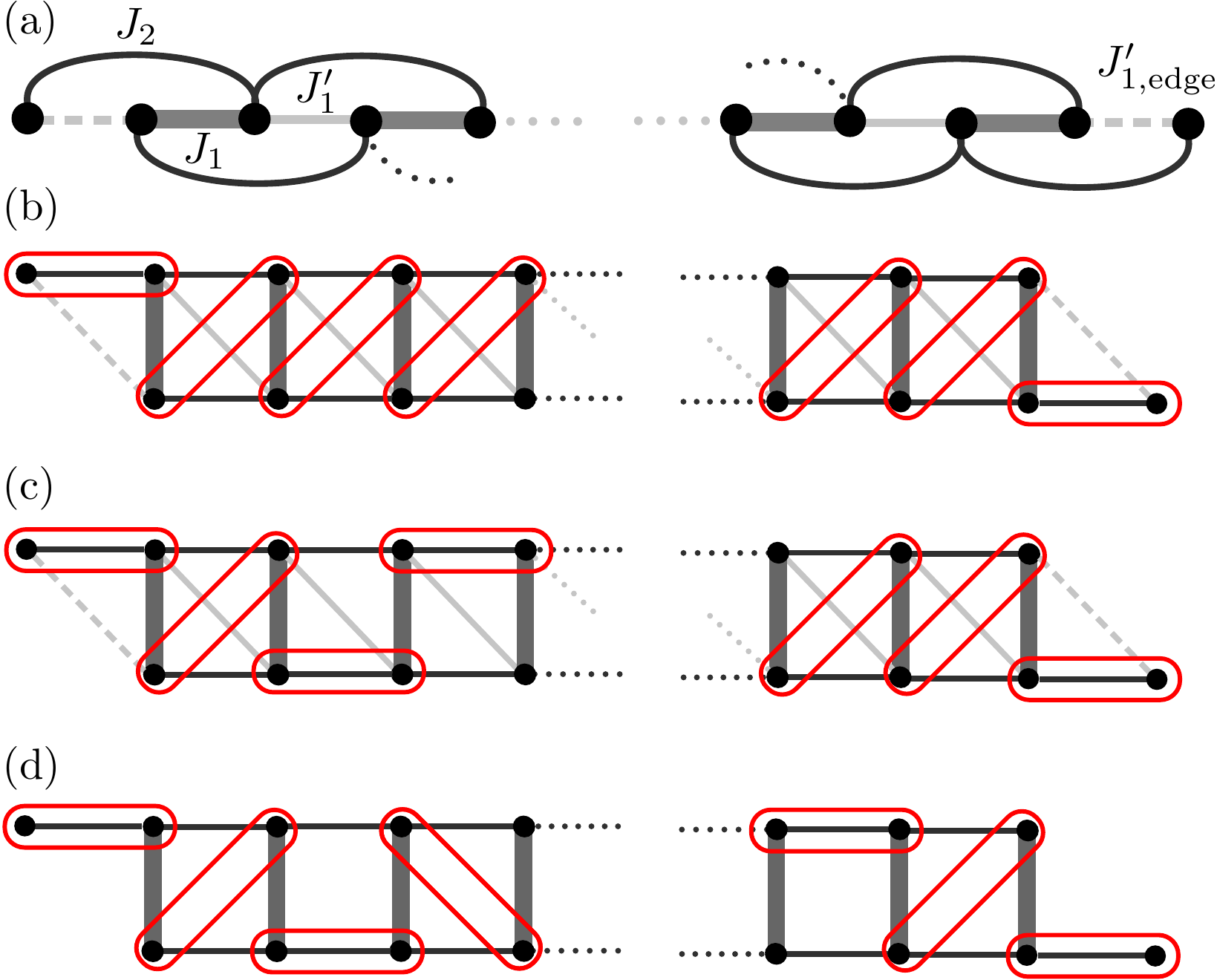}
	\caption{(a) Lattice structure of the $J_1$$-$$J_1'$$-$$J_2$ chain. Snapshots of (b) the $\mathcal{D}_3$-VBS, (c) the mixed-VBS, and (d) symmetry-protected $J_1'=0$ Haldane state, where valence bonds are symbolized by red ellipses and emergent spins-1's are represented by bold lines. The lattices illustrated in (b-d) are topologically equivalent ladder representations of the lattice in (a). The $J_1'$ on both edges (indicated by dashed lines) are set to be zero in our calculations (refer to the main text for further details).}
	\label{fig:lattice}
\end{figure}

The Hamiltonian for the $J_1$-$J_1'$-$J_2$ chain is given by:
\begin{equation}
    H= J_1 \sum_{i=\text{even}} \mathbf S_i \cdot \mathbf{S}_{i+1}  + J_1' \sum_{i=\text{odd}} \mathbf S_i \cdot \mathbf{S}_{i+1} + J_2 \sum_{i} \mathbf S_i \cdot \mathbf{S}_{i+2}
    \label{eq:ham}
\end{equation}
where $\mathbf S_i$ are Heisenberg spin-1/2 operators at site $i$, $J_1$ and $J_1'$ are FM alternating nearest neighbor interactions, and $J_2$ is the AFM next-nearest-neighbor coupling. The lattice of our model is depicted in Fig.~\ref{fig:lattice}. In our analysis, we parametrize the model via the dimerization parameter $\beta=J_1'/J_1$ and the frustration parameter $\alpha=J_2/|J_1|$. 

At $\beta=0$ (Fig.~\ref{fig:lattice}(d)), our system (\ref{eq:ham}) can be effectively mapped onto an $S=1$ Heisenberg chain \cite{Hida1991, Watanabe1993} since two spins coupled by $J_1$, i.e., a rung in the ladder representation, form a spin-triplet pair via three $S^z$ states: 
$\ket{1}_{i,i+1}=\ket{\uparrow}_i\ket{\uparrow}_{i+1}$, 
$\ket{0}_{i,i+1}=(\ket{\uparrow}_i\ket{\downarrow}_{i+1}+\ket{\downarrow}_i\ket{\uparrow}_{i+1})/\sqrt{2}$,
$\ket{-1}_{i,i+1}=\ket{\downarrow}_i\ket{\downarrow}_{i+1}$ for $S^z=1$, $0$, and $-1$ states, respectively. Therefore, the effective model for $\beta=0$ is
\begin{equation}
H_{\rm eff}=\frac{J_2}{2}\sum_i \tilde{\bf S}_i \cdot \tilde{\bf S}_{i+1}-\frac{J_1}{4}L,
\label{effham}
\end{equation}
where $\tilde{\bf S}_i$ is a spin-1 operator defined as a resultant spin $\tilde{\bf S}_i={\bf S}_{2i}+ {\bf S}_{2i+1}$.

We study finite-size systems with lengths up to $L=600$ sites using the DMRG method. Open boundary conditions (OBC) are applied unless specified otherwise. As confirmed below, our system is marked by a Haldane-like VBS state throughout the entire range of the gapped phase. Hence, following Ref.~\onlinecite{Agrapidis2019}, we explicitly control the edge states of our system by imposing $J_1'^{\text{edge}}=0$ at both chain edges \cite{Oshikawa1992,Kennedy1992a,Kennedy1992b,Kennedy1990,Pollmann2012}. This effectively eliminates the degeneracy of edge spins and facilitates the correct computation of the {\it bulk} spin gap (see also App. \ref{App:boundaries}). We keep up to $m=5500$ density-matrix eigenstates in the renormalization procedure, resulting in an error $\epsilon/L\sim10^{-8}$ for fixed system size. Consequently, we can perform a reliable finite-size scaling analysis extrapolating to the thermodynamic limit.

\section{Results}

\subsection{Ground-state phase diagram}

It has been established that our system (\ref{eq:ham}) is in
a FM phase for $\alpha<\beta/[2(1+\beta)]$ and in a singlet
gapped phase for $\alpha>\beta/[2(1+\beta)]$ \cite{Agrapidis2017}.
However, the magnetic structure of the gapped phase has not
been determined yet except at $\beta=1$ and $\beta=0$. When
$\beta=1$, the ground state is a VBS state with valence bond
formation between third-neighbor spin-1/2's due to the
order-by-disorder mechanism, leading to a spontaneous breaking of
the translational symmetry [Fig.~\ref{fig:lattice}(b)] \cite{Agrapidis2019}. This specific state is designated as the
$\mathcal{D}_3$-VBS state. Whereas at $\beta=0$, the ground
state is interpreted as the so-called (symmetry-protected) Haldane
state since the system is effectively mapped onto a spin-1
Heisenberg chain~\eqref{effham}
\cite{Hida1991,Watanabe1993,Gu2009,Tasaki2020}.
This state translates into the formation of a valence bond on
either the diagonal pair or $J_2$ bond in the ladder representation
of our system [Fig.~\ref{fig:lattice}(d)]. This state is referred
as the mixed-VBS state in the sense of coexisting different types
of valence bonds.

\begin{figure}[t]
    \centering
    \includegraphics[width=\columnwidth]{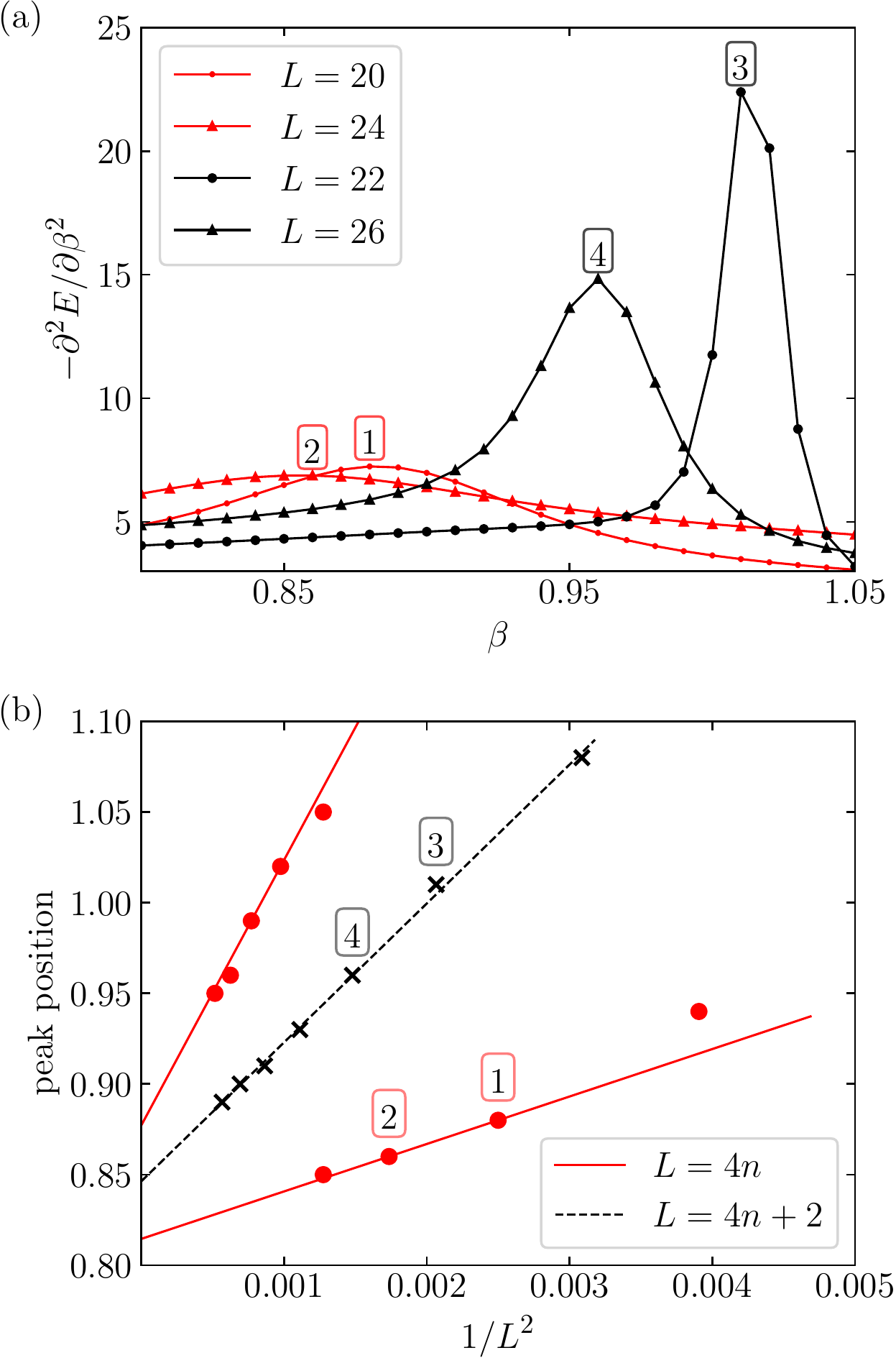}
    \caption{(a) Example of second partial derivative of the ground-state energy as a function of $\beta$ for fixed $\alpha=0.42$. (b) Examples of finite-size scaling analysis for the peak positions in the second derivative. The linear scaling is achieved by plotting the peak positions as a function of $1/L^2$. The numbers enclosed within squares correspond to the peaks in (a). The presence of two red lines is due to the multiple peak feature by finite-size effect (see text).
    }
    \label{fig:Peaks}
\end{figure}

Our findings indicate that the gapped region within the range $0<\beta<1$ is predominantly characterized by the $\mathcal{D}_3$-VBS or the mixed-VBS phase. These are considered as natural extensions of the limiting cases at $\beta=1$ and $\beta=0$. Note that, unlike for the $\beta=0$ limit, only one direction of the diagonal valence bond is allowed in the mixed-VBS state at $\beta>0$
because the mirror symmetry of the ladder is broken when $J_1'$
is finite [Fig.~\ref{fig:lattice}(c)].

We have determined the phase boundary through detection of peaks in the second derivative of the ground-state energy. For this analysis, periodic chains with up to 48 sites are used. In Fig. \ref{fig:Peaks}(a) we plot the second derivative as a function of $\beta$. Even though a discernible peak at a specific $\beta$ value for a fixed system size is noticeable, the peak position seems to vary depending on the system size. Additionally, multiple peaks appear in the parameter region adjacent to the FM instability when plotted against $\beta$ (not shown here). This multiple peak feature is typically interpreted as finite-size effects arising from discrete quantum numbers relative to the system size. Nevertheless, we can achieve a reasonable extrapolation of the peak positions to a {\it singular} value in the thermodynamic limit, by performing a finite-size scaling analysis of the second-derivative peaks using a linear function as illustrated in Fig. \ref{fig:Peaks}(b). Note that in this case there are two scaling lines for $L=4n$ due to the multiple peak feature. Furthermore, a more accurate convergence of extrapolated values in the thermodynamic limit may be achieved if we employ a polynomial function instead of a linear one. We also find that other than the above peaks we see a system-size independent single peak indicating the FM and $\mathcal{D}_3$-VBS phase boundary.

\begin{figure}[t]
    \centering
    \includegraphics[width=\columnwidth]{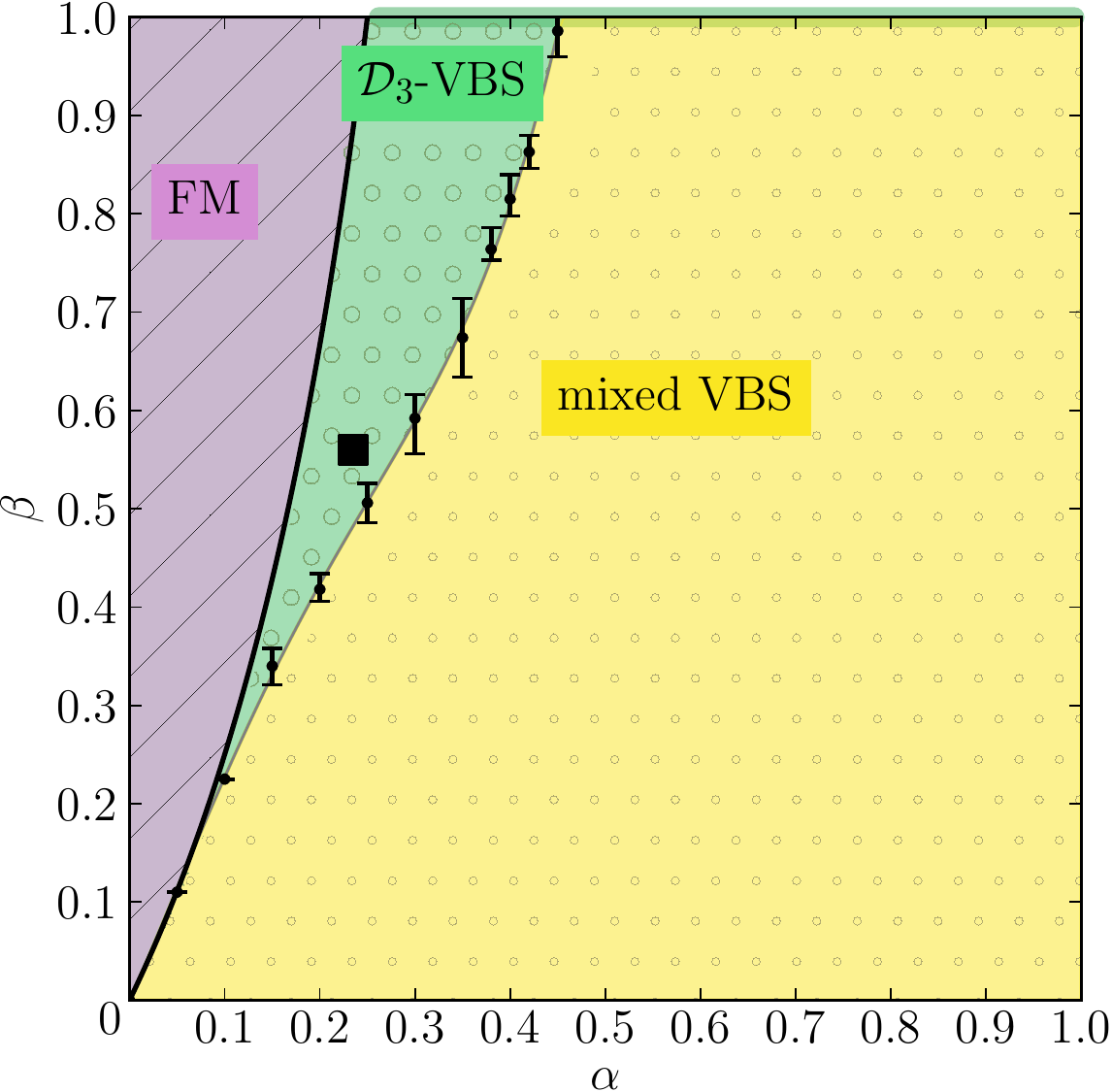}
    \caption{Ground-state phase diagram of the $J_1$-$J_1'$-$J_2$ chain in the $\alpha$-$\beta$ plane. The FM phase line was determined via spin-wave theory in Ref.~\onlinecite{Agrapidis2017}. The boundary between the $\mathcal{D}_3$-VBS and the mixed-VBS phases is determined via the peak position of the second derivative of the ground-state energy. At $\beta=1$ the system is in the $\mathcal{D}_3$-VBS phase for $\alpha>1/4$, as indicated by the green line. The black square indicates the parameter set of LiCuSbO$_4$.}
    \label{fig:PD}
\end{figure}

The resultant $\alpha$-$\beta$ phase diagram is shown in Fig.~\ref{fig:PD}, where the width of the error bar is estimated based on the variations in the extrapolated values derived from the preceding finite-size scaling analysis. The green and yellow areas correspond to the $\mathcal{D}_3$-VBS and the mixed-VBS phases, respectively. Note that the gapped phase at $\beta=1$ is in the $\mathcal{D}_3$-VBS at any $\alpha (>1/4)$ \cite{Agrapidis2019}. The two VBS phases can be generally distinguished based on the ratio of spin-spin correlations between the second- and third-neighbor spin-1/2's. Specifically, the system is typified by the $\mathcal{D}_3$-VBS phase (or the mixed-VBS phase) when the valence bond between the third neighbors is more robust (or weaker) than that between the second neighbors (see Sec. \ref{ex_two_VBS_ratio}). However, this trend is less obvious as $\beta$ decreases, and the strength of the second- and third-neighbor valence bonds becomes equivalent in the limit of $\alpha=\beta=0$. An additional intriguing aspect is the identification of a hidden AFM order validated by a finite SOP for both VBS states \cite{Oshikawa1992} (see Sec. \ref{subsec:sop}). Consequently, both VBS states can be characterized as Haldane-like VBS states. This is further confirmed by the analysis of their entanglement spectra (ES) [see Sec.~\ref{subsec:ES}].

\subsection{Spin gap}\label{subsec:spingap}

\begin{figure}[t]
    \centering
    \includegraphics[width=\columnwidth]{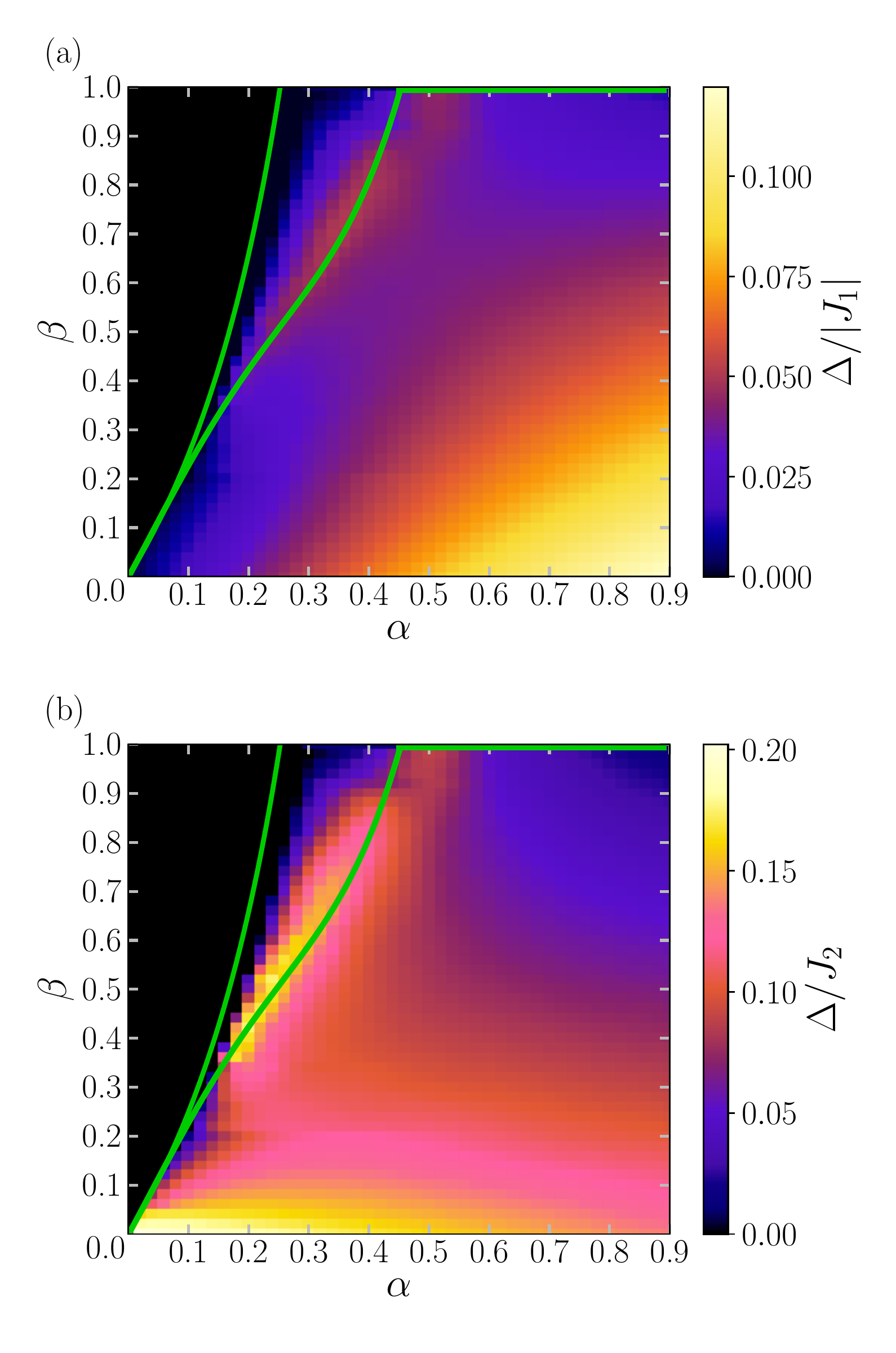} 
    \caption{Spin gap in the $\alpha$-$\beta$ plane in terms of (a) $|J_1$| and (b) $J_2$. Color bars shows the magnitude of the gap. The green lines denote the boundaries between the FM, the $\mathcal{D}_3$-VBS, and the mixed-VBS phases.
    }
    \label{fig:SpinGap}
\end{figure}

We start by computing the spin gap as a function of $\alpha$ and $\beta$. Since the whole gapped region is characterized by a Haldane-like VBS state, we specifically eliminate the ground-state degeneracy due to the edge states by setting $J'^{1,\mathrm{edge}}_1=0$ at both ends of the chain \cite{Kennedy1990,Agrapidis2019,Oshikawa1992,Kennedy1992a}. By evaluating the ground-state energy $E_0(S^z=0)/L$ in the thermodynamic limit, we have confirmed that this choice of edge structure does not affect the bulk ground-state properties (see Appendix~\ref{App:boundaries}). With the degeneracy of edge spins removed, we can determine the spin gap as the energy difference between the singlet ground state and the first excited triplet state:
\begin{align}\label{eq:spingap}
    &\Delta = \lim_{L\to\infty} \Delta(L), \nonumber \\
    &\Delta(L) = E_0(L, S^z=1)-E_0(L, S^z=0),
\end{align}
where $L$ is the susyem size and $E_0(L, S^z)$ is the ground-state energy in the $S^z$ spin sector for a fixed system size $L$. Examples of finite-size scaling for $\Delta(L)$ are shown in Appendix~\ref{App:FSS}. The extrapolated spin gap $\Delta$ in the thermodynamic limit is plotted in Fig.~\ref{fig:SpinGap}. As the mapping to the spin-1 model (\ref{effham}) is the most efficacious at $\beta=0$, the gap is expected to increase with approaching $\beta=0$. 

Let us first see the gap at $\beta=0$. When we consider the gap in terms of $|J_1|$, it increases almost linearly with $\alpha$, reflecting the effective exchange coupling $J_2/2$ in (\ref{effham}). On the other hand, $\Delta/J_2$ increases slowly with decreasing $\alpha$. This is because the mapping to (\ref{effham}) is generally good for any $\beta=0$ but is exact only in the limit of $\alpha=\beta=0$. Consequently, the gap saturates at $\Delta/J_2=0.2045$, which is that for the simple $S=1$ Heisenberg chain with AFM exchange interaction $J_2/2$, with approaching $\alpha=\beta=0$ \cite{Agrapidis2017}. Given this, one might naively anticipate the gap to reduce with increasing $\beta$ due to the potential obstruction of the effective spin-1 formation. However, it is worth noting that the gap has a maximum around the boundary between the $\mathcal{D}_3$-VBS and the mixed-VBS phases with fixed $\beta$. This behavior is more evident when taking the gap in units of $J_2$ [Fig.~\ref{fig:SpinGap}(b)] than in unit of $|J_1|$ [Fig.~\ref{fig:SpinGap}(a)]. Nevertheless, since the peak is relatively sharp, its position remains nearly unaffected by the energy unit $|J_1|$ or $J_2$. This suggests that frustration is maximized around the phase boundary where the two VBS phases highly compete. The emergence of this peak significantly validates the existence of a phase transition.

We should also note that the present gap, particularly within the small $\alpha$ and large $\beta$ region, is somewhat larger than the one previously estimated in Ref.~\onlinecite{Agrapidis2017}. In other words, the gap was previously underestimated without explicit lifting the degeneracy of edge spins, meaning the free spin state at the system edges within the VBS structure was not adequately defined. This validates that the $\mathcal{D}_3$-VBS is a kind of Haldane state.

\subsection{String order parameter}\label{subsec:sop}

\begin{figure}[t]
    \centering
    \includegraphics[width=\columnwidth]{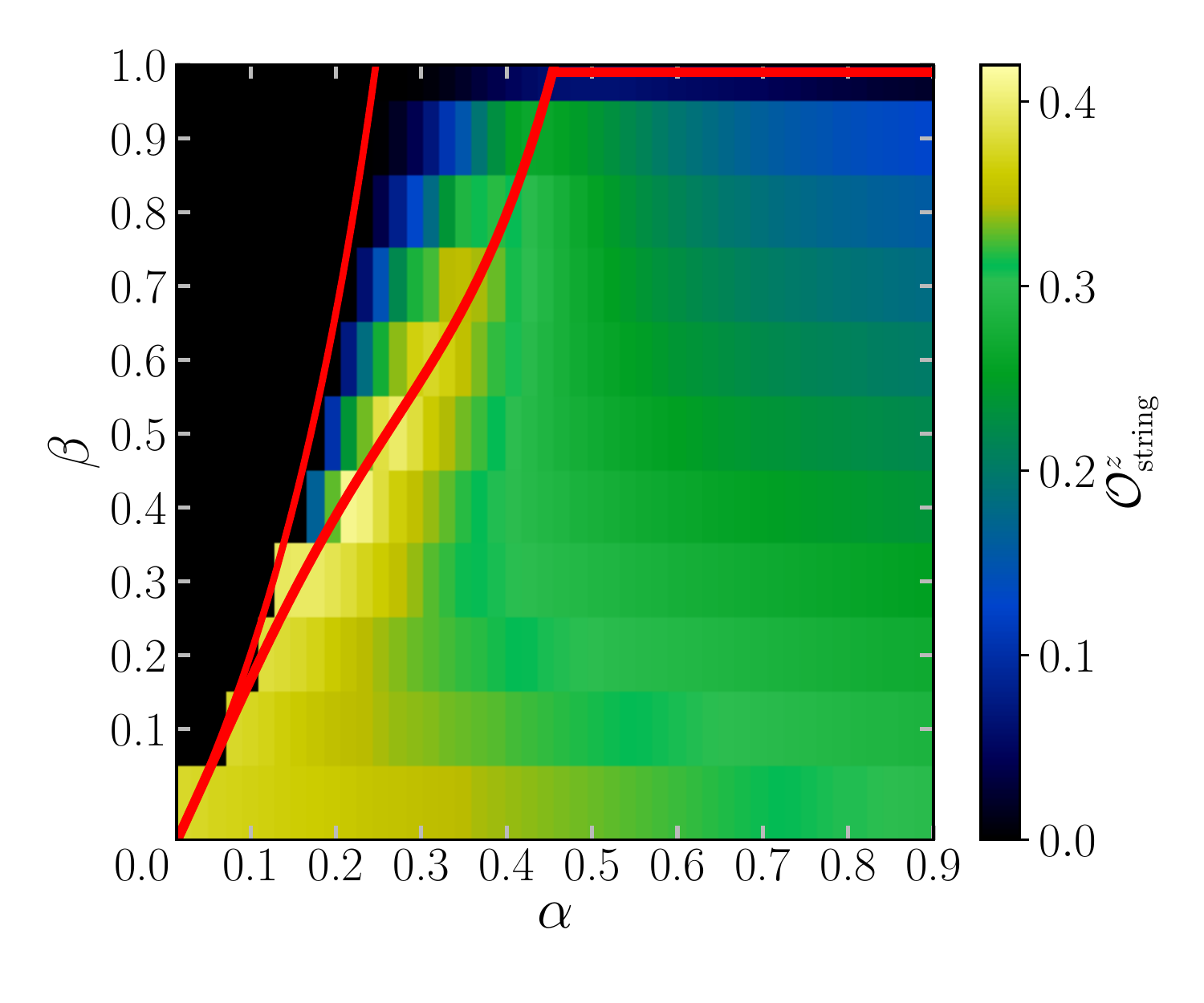}
    \caption{String order parameter $\langle \mathcal O^z_\mathrm{string} \rangle$ in the $\alpha$-$\beta$ plane.  The red lines denote the  boundaries between the FM, the $\mathcal{D}_3$-VBS, and the mixed-VBS phases.
    }
    \label{fig:SOP}
\end{figure}

As discussed above, it appears that the entire gapped phase may be characterized by a Haldane-type VBS state, which is topologically nontrivial. Thus, we proceed to calculate the SOP which serves a good indicator to evaluate the the stability of such a VBS state. This quantity was originally defined for spin-1 systems \cite{denNijs1989}. However, for spin-1/2 systems, the SOP can be suitably redefined as \cite{Agrapidis2019,Kohmoto1992,Hida1992}
\begin{align}
    \mathcal O^z_\mathrm{string} =& \lim_{|j-k|\to\infty} (-1)^\frac{j-k-2}{2} \langle (S^z_k +S^z_{k+1}) \\ \nonumber
    &\prod_{l=k+2}^{j-1} S^z_l(S^z_j+S^z_{j+1})\rangle,
    \label{eq:SOP}
\end{align}
where $S^z_j$ is the $z$-component of the spin-1/2 at site $j$. In our model \eqref{eq:ham}, the dimerization introduced by $J_1'<J_1$ breaks the translational symmetry of the chain. Hence, our system has a stronger FM coupling between the even-odd sites than between the odd-even sites. This leads to the emergence of effective spins-1 degrees of freedom on each $J_1$ bond. Therefore, $j$ and $k$ in Eq.~\eqref{eq:SOP} must be even. Indeed, we have also confirmed that if $j$ and $k$ are taken to be odd, the value of $\mathcal O^z_\mathrm{string}$ is always extrapolated to zero in the thermodynamic limit. Since OBC are applied here, we take $k=L/4$ and $j=3L/4$ to perform a systematic finite-size scaling with sufficiently minimizing the influence of Friedel oscillations.

The extrapolated values of the SOP are plotted in Fig.~\ref{fig:SOP}. We find that the SOP is finite throughout the gapped region of the phase diagram. Moreover, the trend of $\mathcal O^z_\mathrm{string}$ in the $\alpha$-$\beta$ space is very similar to that of $\Delta/J_2$ as seen in Fig.~\ref{fig:SpinGap}(b). Specifically, it diminishes when moving from the bottom to the top right corner, and it has relatively large value around the boundary of the two VBS phases.
Intriguingly, the SOP around the phase boundary can exceed $\mathcal O^z_\mathrm{string}=0.3743$ for the $S=1$ Heisenberg chain~\cite{White1993}: For instance, at $(\alpha,\beta)=(0.2,0.4)$ $\mathcal O^z_\mathrm{string}$ is $0.4101$ which is even closer to $\mathcal O^z_\mathrm{string}=4/9=0.4444\cdots$ for the perfect VBS state of the AKLT model \cite{Affleck1987}. This finding illustrates a possible enhancement of valence bond formation by frustration. This enhancement of the SOP at the phase boundary is reflected as a minimization of the lowest level in the ES (see Sec.~\ref{subsec:ES}).

\subsection{Existence of two distinct VBS states}\label{ex_two_VBS_ratio}

Thus far, the existence of a Haldane-type VBS state has been confirmed in the whole gapped region of the $\alpha$-$\beta$ parameter space. However, the exact arrangement of the valence-bond formation remains to be elucidated. It is known that our system is in a Haldane mixed-VBS state when $\beta=0$, wherein the valence bonds are formed along both the $J_2$ couplings and the diagonal directions of the ladder representation [Fig.~\ref{fig:lattice}(d)]. On the other hand, when $\beta=1$, our system exhibits a unique $\mathcal D_3$-VBS state, characterized by valence bonds forming exclusively along third neighbors in a direction dictated by spontaneous FM dimerization \cite{Agrapidis2019}. What remains to be determined is how these two VBS phases develop in the parameter region $0<\beta<1$, and whether there exists a possible additional VBS phase in any region of the $\alpha$-$\beta$ plane.

\begin{figure}[t]
    \centering
    \includegraphics[width=\columnwidth]{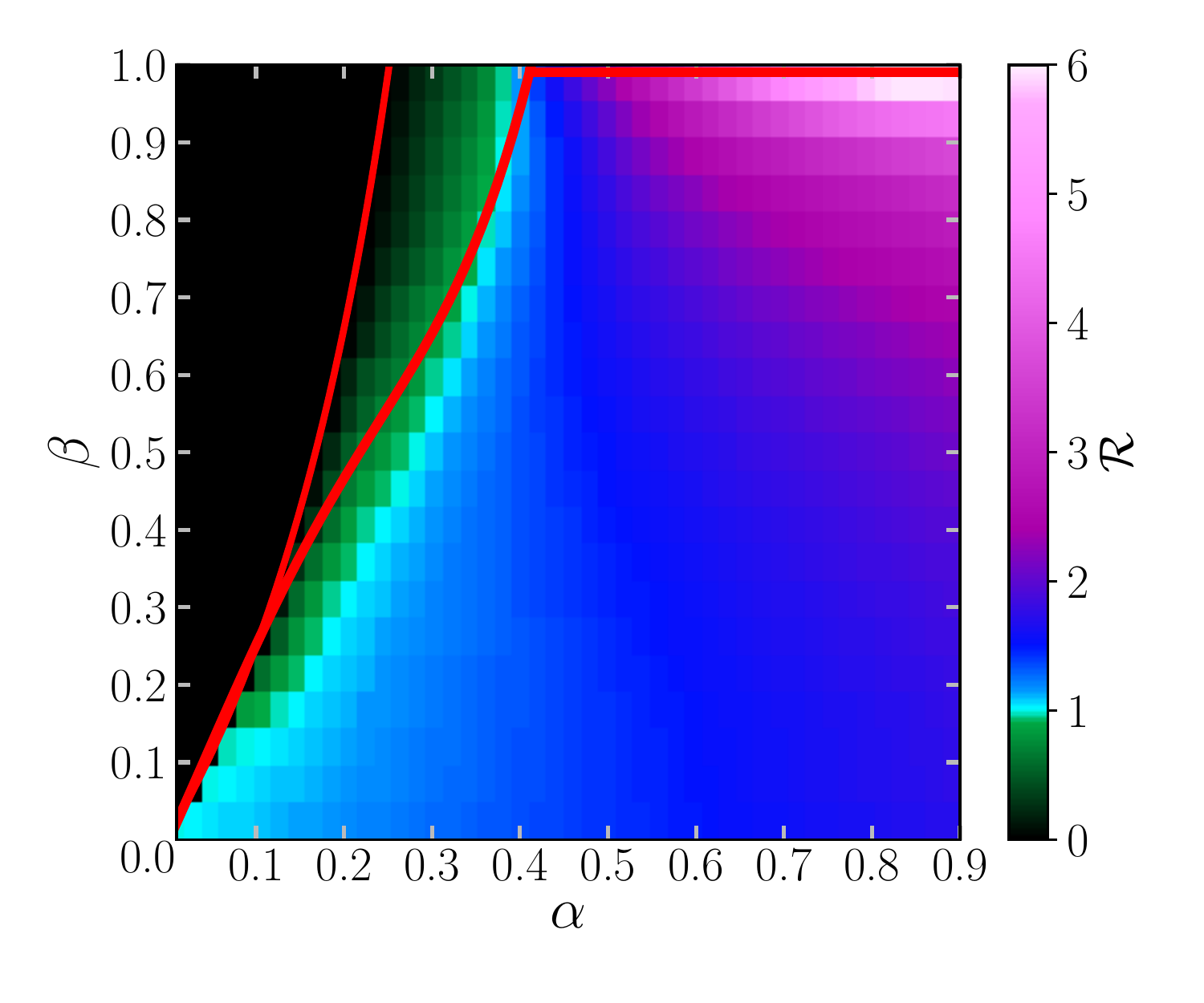}
    \caption{Ratio between the second- and third-neighbor spin-spin correlations in the $\alpha$-$\beta$ plane. The red lines denote the boundaries between the FM, the $\mathcal{D}_3$-VBS, and the mixed-VBS phases.}
    \label{fig:ratio}
\end{figure}

Given that we have verified that the SOP (with assuming only $J_1$ rungs to exhibit spin-1 degrees of freedom) is finite within the whole gapped region, all valence bonds are likely to be formed between neighboring $J_1$ rungs. Considering the ladder representation, a connection between neighboring $J_1$ rungs involves the nearest-, second-, and third-neighbor bonds in the original chain. However, the formation of nearest-neighbor valence bonds is disregarded except for $\beta=0$ as it corresponds to a FM $J_1'$ coupling. Therefore, we can focus only on the possibility of the second- and/or third-neighbor valence bond formations. A simple approach to determine the structure of a VBS state is to calculate the dimerization order parameter defined by $\mathcal D_\delta=\lim_{L\to\infty}|\langle \mathbf S_{i-\delta} \cdot \mathbf S_{i} \rangle - \langle \mathbf S_{i} \cdot \mathbf S_{i+\delta} \rangle|$ where $\delta$ assigns the length of the considered valence bond. Since the dimerization between $J_1$ and $J_1'$ is explicitly introduced in our system \eqref{eq:ham} for $0\le\beta<1$, $\mathcal D_\delta$ is always finite for odd $\delta$. This order parameter can also detect spontaneous dimerization. We thus calculate $\mathcal D_2$ for the second-neighbor bonds and find it to be always zero in the thermodynamic limit (not shown). This suggests that a VBS state composed solely of second-neighbor valence bond formations does not exist. Accordingly, we conclude that the gapped region at $0<\beta<1$ is covered by either the $\mathcal{D}_3$-VBS or the mixed-VBS phases.

To ascertain how the two VBS phases are distributed within the $\alpha$-$\beta$ plane, we evaluate the relative strength of second- and third-neighbor spin-spin correlations. We define the ratio as:
\begin{align}
  \mathcal R &= \lim_{L\to\infty}\mathcal R(L), \nonumber\\
  \mathcal R(L) & = \left | \frac{\langle \mathbf S_{L/2}\cdot \mathbf S_{L/2 +2}\rangle}{\langle \mathbf S_{L/2}\cdot\mathbf S_{L/2+3}\rangle} \right|.
\end{align}
This is measured at the center of the chain to minimise the effect of Friedel oscillations. In general, a larger correlation does not always imply valence bond formation. Nevertheless, in this context, the relative strength $\mathcal R$ should provide a good indication as to the transfer between the $\mathcal{D}_3$-VBS and the mixed-VBS phases since it represents a contiguous change between the leg and the diagonal valence bonds within each plaquette in the ladder representation, rather than a complete redistribution of valence bonds.

When $\mathcal R >1$ ($\mathcal R <1$) the AFM spin-spin correlation between second-neighbor (third-neighbor) spins exceeds that between third-neighbor (second-neighbor) spins. Note that both second- and third-neighbor spin-spin correlations are always AFM in the gapped region. Intuitively, a spin-singlet formation on a bond with a larger AFM spin-spin correlation is energetically favorable. Hence, we may expect our system to be in the $\mathcal{D}_3$-VBS state for $\mathcal R <1$ and in the mixed-VBS state for $\mathcal R >1$. Our results for $\mathcal R$ are plotted in Fig.~\ref{fig:ratio}. It exhibits a similar behavior to that of the gap $\Delta/J_2$ [Fig.~\ref{fig:SpinGap}(b)] and of the SOP [Fig.~\ref{fig:SOP}]. Indeed, we observe $\mathcal R \sim 1$ at the boundary between the two VBS phases, indicating the competition between second- and third-neighbor valence bonds. This seems to reasonably explain the phase transition for $0<\beta<1$.

It should be noted, however, that this approach is not universally reliable in detecting the valence bonds structure. For example, at $\beta=1$, $\mathcal R$ is larger than 1 and simply increases with $\alpha$ within the $\mathcal{D}_3$-VBS phase. For larger values of $\alpha$, our system can be regarded as weakly coupled AFM chains. In such cases, avoiding the formation of the second-neighbor valence bonds could be possible due to large fluctuation. Even so, an unresolved issue remains: Do the physical quantities asymptotically approach the undimerized limit $\beta=1$, or does this limit serve as a singular point? We believe that insights into this question could be gained from an examination of the Berry phase in the $\alpha$-$\beta$ plane, specifically near the $\beta=1$ line. Nonetheless, this goes beyond the scope of our current work.


\subsection{Entanglement Spectra} \label{subsec:ES}
\begin{figure}[t]
    \centering
    \includegraphics[width=0.9\columnwidth]{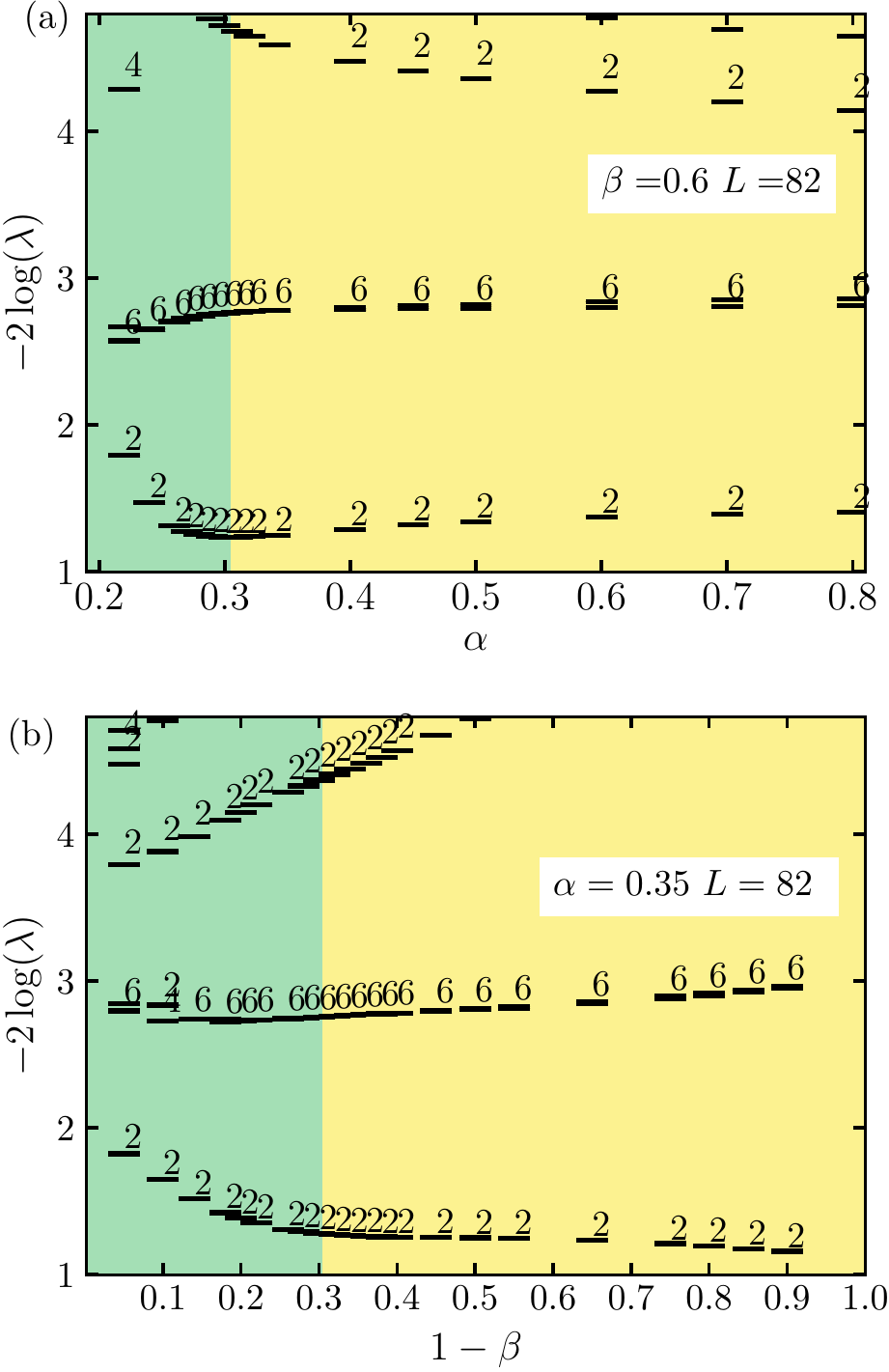}
    \caption{Plots of the entanglement spectrum (ES) of a system with PBC and size $L=82$ (a) as a function of $\alpha$ for $\beta=0.6$, (b) as a function of $1-\beta$ for $\alpha=0.35$. The green region denotes $\mathcal{D}_3$-VBS, while the yellow denotes the mixed-VBS state. Numbers indicate the degeneracy of the state.}
    \label{fig:ESalpha_beta_part}
\end{figure}

Entanglement related quantities are often used to study topological properties of VBS states including the Haldane state \cite{Pollmann2010}. Having confirmed that both VBS states in our phase diagram are classified as of the Haldane-type, it is interesting to confirm their topological nature by looking at the low-lying eigenstates of the ES \cite{Li2008a,Fidkowski2010,Li2008b,Vidal2003}. In our calculations, we consider the ES obtained by partitioning the system into two subsystems $A$ and $B$. Defining $\xi_{\lambda}$ in the Schmidt decomposition of the ground state $\ket{\psi}$ as
\begin{align}
\ket{\psi}=\sum_{\lambda}e^{-\xi_{\lambda}/2}
\ket{\lambda}_{A}\ket{\lambda}_{B} ,
\end{align}
where $\ket{\lambda}_{A}$ ($\ket{\lambda}_{B}$) is the orthonormal basis for the subsystem $A$ ($B$), we can interpret the ES as the energy spectrum of the entanglement Hamiltonian $\mathcal{H}_{e}$ defined as
$e^{-\mathcal{H}_{e}}=\rho_{A}=\mathrm{Tr}_{B}\ket{\psi}\bra{\psi}
=\sum_{\lambda}e^{-\xi_{\lambda}}\ket{\lambda}_{A}\bra{\lambda}_{A}$.  
If the system size is sufficiently larger than the correlation length, the ES can be described by the two virtual edge states; the ES therefore represents the gapless edge modes in its lowest-lying eigenstate. Since spin-spin correlations decays exponentially in the VBS phase, the correlation length is expected to be quite short.

We now consider periodic chains with system size $L=4n+2=2(2n+1)$ and divide them in half. Since the two subsystems consist of an odd number of sites, we can directly detect the emergent spin-1/2 edge states, if they exist. Our results for the ES with size $L=82$ are shown in Fig.~\ref{fig:ESalpha_beta_part}. We clearly see 2-fold degeneracy of the lowest-lying eigenstate. In our dimerized system, each of the subsystems contains only one free spin-1/2 at one edge. Accordingly, non-entangled two spin-1/2's with total $S^z=0$ lead to the observed 2-fold degeneracy. Since this picture is valid for both the $\mathcal{D}_3$-VBS and the mixed-VBS states, the 2-fold degeneracy is unchanged through the phase boundary. It is worth mentioning that, as seen in Fig.~\ref{fig:ESalpha_beta_part}(a), the lowest-lying level in the ES reaches a minimum around the phase boundary. This is a reflection of the stability of the VBS state as validated by the SOP. Moreover, this supports the validity of the determined VBS structures. We also find that the second eigenstate is 6-fold degenerate, with four states located in the $S^z=0$ sector and the remaining two at $S^z=\pm1$. More details are discussed in App. \ref{appendix:ES}.

Note that the $\mathcal{D}_3$-VBS state exhibits 4-fold degeneracy in the lowest-lying eigenstate for the undimerized case $\beta=1$ \cite{Agrapidis2019}. This is because there are two possible ground states with spontaneous FM dimerization. As a result, the free spin-1/2 can be positioned at both edges of each subsystem. The presence of this 2-fold degeneracy at both cut edges cumulatively leads to the 4-fold degeneracy. More details are given in App. \ref{appendix:ES}.

\subsection{Evolution of the gap with explicit dimerization}\label{subsec:Gap_exp_dim_J3}

\begin{figure}[t]
    \centering
    \includegraphics[width=0.9\columnwidth]{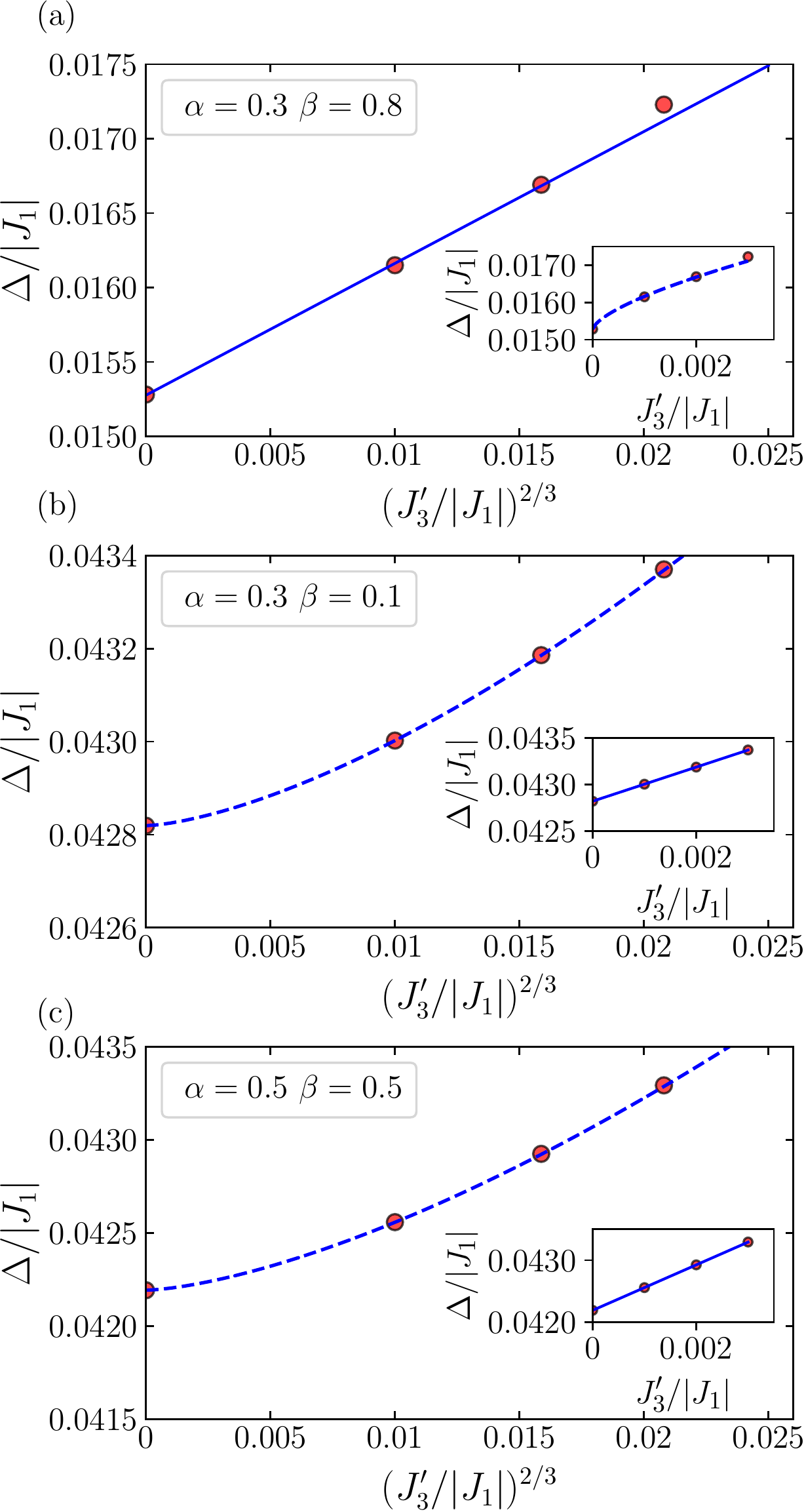}   
    \caption{Spin gap as function of $(J_3'/|J_1|)^{2/3}$ for (a) $(\alpha,\beta)=(0.3, 0.8)$, (b) $(0.3, 0.1)$, and (c) $(0.5, 0.5)$. Inset: The same quantity as a function of $J_3'/|J_1|$. Solid lines are linear fits and dotted lines are guides for eyes.
    } 
    \label{fig:gapJ3}
\end{figure}

Based on the above results, we have found that the gapped region in the ground-state phase diagram is divided into two kinds of VBS phases, namely, the $\mathcal{D}_3$-VBS and the mixed-VBS phases. Nevertheless, the $\mathcal{D}_3$-VBS phase is confined to only a narrow range near the FM instability except for the $\beta=1$ line. To substantiate this observation, following Ref.~\onlinecite{Agrapidis2019}, we explore the dependence of the spin gap on an additional AFM third-neighbor exchange interaction $J_3'$. The modified Hamiltonian is written as
\begin{equation}
    H'=H+J_3'\sum_{i=\mathrm{even}} \mathbf S_i \cdot \mathbf{S}_{i+3},
\end{equation}\label{eq:hamJ3}
where $J_3'(>0)$ is added only for even $i$ in Fig.~\ref{fig:lattice}. This third-neighbor bond is associated with the valence bond in the $\mathcal{D}_3$-VBS state. If the system is in the $\mathcal{D}_3$-VBS state which is a kind of spin-Peierls state, the spin gap is expected to increase as $\Delta-\Delta(J_3'=0) \propto J_3^{\prime\frac{2}{3}}$ at small $J_3^\prime$ \cite{Uhrig1999}. 

In Fig.~\ref{fig:gapJ3}(a) we plot the spin gap as a function of $(J_3'/|J_1|)^{2/3}$ for the $\mathcal{D}_3$-VBS state at $(\alpha, \beta)=(0.3, 0.8)$. We can see a power-law behavior of the spin gap with exponent $2/3$ at $0<J_3'\lesssim0.002$. We thus confirm that the $\mathcal{D}_3$-VBS state can exist even if our system is away from $\beta=1$. On the other hand, the gap behavior clearly deviates from $\Delta-\Delta(J_3'=0) \propto J_3^{\prime\frac{2}{3}}$ for the mixed-VBS states at $(\alpha,\beta)=(0.3, 0.1)$ and $(0.5, 0.5)$, as shown in Fig.~\ref{fig:gapJ3}(b,c). Note that these $(\alpha,\beta)$ points are in the mixed-VBS phase, but not very far from the boundary to the $\mathcal{D}_3$-VBS phase. For these points, the spin gap increases linearly with $J_3^\prime$, i.e., $\Delta-\Delta(J_3'=0) \propto J_3^{\prime}$. This implies that the lowest spin excitation is given by the collapse of a spin-singlet pair on the second-neighbor bond. This is because the second-neighbor bond involves a direct coupling by AFM $J_2$ as well as an indirect coupling via FM $J_1$ and AFM $J_3^{\prime}$, i.e, a superexchange path through $J_1$ and $J_3^{\prime}$, so that the second-neighbor exchange interaction is effectively linearly enhanced by $J_3^{\prime}$. Morevoer, this also means that the valence bond between the third-neighbor spins is stronger than that between the second-neighbor spins in the mixed-VBS phase, and their strengths become equivalent in the limit of $\alpha=\beta=0$.

Furthermore, it is interesting to consider the difference value $\delta \Delta/\delta J_3^{\prime}$: We obtain $0.0367$, $0.3633$, and $0.8757$ for $(\alpha,\beta)=(0.3, 0.1)$, $(0.5, 0.5)$, and $(0.3, 0.8)$, respectively. Since the valence bonds are formed only on the $J_3^{\prime}$ bonds in the $\mathcal{D}_3$-VBS phase, the increase of $J_3^{\prime}$ directly contributes to the gap increase. The different values are obtained via the gap increase from $J_3^{\prime}=0$ to $0.001$. More detailed data is shown in App.~\ref{App:GapJ3}.

\section{Summary and discussion}

We studied the frustrated dimerized FM-AFM $J_1$-$J_1'$-$J_2$ chain using the DMRG method. We obtained the ground-state phase diagram for a wide range of $J_1'/J_1$ and $J_2/|J_1|$  based on the results for the second derivative of energy, spin gap, spin-spin correlations, SOP, and ES. The phase diagram consists of a FM and two gapped VBS phases. These VBS phases are characterized as the $\mathcal{D}_3$-VBS and the mixed-VBS phase. In the $\mathcal{D}_3$-VBS state, valence bonds are formed between third-neighbor spin-1/2's. The mixed-VBS state, on the other hand, exhibits a coexistence of both second- and third-neighbor valence bonds. Remarkably, these VBS states are identified as being of the Haldane-type, characterized by a finite value of the SOP and the 2-fold degeneracy of the ES. It is remarkable that the formation of valence bonds is significantly stabilized at the boundary between the two VBS phases. Moreover, at this boundary, the gap and SOP are maximized, while the lowest-lying eigenstate of the ES is minimized. These observations underscore the critical role of frustration in the emergence of topological states in our system.

Finally, we briefly discuss the relevance to the experimental observations for LiCuSbO$_4$. For this compound, density-functional theory (DFT) calculations have estimated the effective exchange parameters to be $J_1\approx -160$K, $J'_1\approx 90$K, and $J_2\approx 37.6$K \cite{Grafe2017}. This set of parameters corresponds to $(\alpha,\beta)=(0.235,0.5625)$ in our notation, putting this material in the $\mathcal{D}_3$-VBS phase in our ground-state phase diagram as shown in Fig.~\ref{fig:PD}. Intriguingly, the possibility of exotic phenomena, such as multipolar states in the presence of a magnetic field, has been experimentally suggested \cite{Grafe2017,Bosioi2017}. The presence of multipolar physics in 1D systems under an applied field  has been discussed only within the uniform $J_1$-$J_2$ chain case ($\beta=1)$\cite{Hikihara2008,Sudan2009,Mourigal2012}. We thus believe that the effect of dimerization on the multipolar physics is a question of interest and should be the subject of future studies.

The present study represents the first demonstration of a nontrivial quantum phase transition between topological VBS states in spin-1/2 chains, though the possibility of phase transitions through a complete reconstruction of valence bond structure have been suggested in 2D systems \cite{Vishwanath2004,Slagle2014,DaLiao2022}, 1D systems with alternating different spins \cite{Tonegawa1998,Li2006,Verssimo2023}, and higher spin systems \cite{Kitazawa1997,Miyakoshi2016}. It is very exciting that quasi-1D materials such as LiCuSbO$_4$ [$(\alpha,\beta)=(0.235,0.5625)$] and Rb$_2$Cu$_2$Mo$_3$O$_{12}$ [$(\alpha,\beta)=(0.37,0.65)$] \cite{Hase2004,Agrapidis2017} may be situated near the boundary between two topological phases, and that transitions between topological ordered states could potentially be observed experimentally through controls such as pressure and stretching.

\section*{Acknowledgements}

We thank Krzysztof Wohlfeld for fruitful discussion. This work was supported by Narodowe Centrum Nauki (NCN, Poland) under Project Nos. 2021/40/C/ST3/00177. This research was carried out with the support of the Interdisciplinary Center for Mathematical and Computational Modeling at the University of Warsaw (ICM UW) under grants No. G81-4 and No. G93-1613. This project is funded by the German Research Foundation (DFG) via the projects A05 of the Collaborative Research Center SFB 1143 (project-id 247310070). The code to reproduce the data and figures presented in this manuscript is available at Ref. \cite{zenodo}.

\appendix


\section{Special open boundary conditions} \label{App:boundaries}

\begin{figure}[tbh]
	\centering
	\includegraphics[width=0.8\columnwidth]{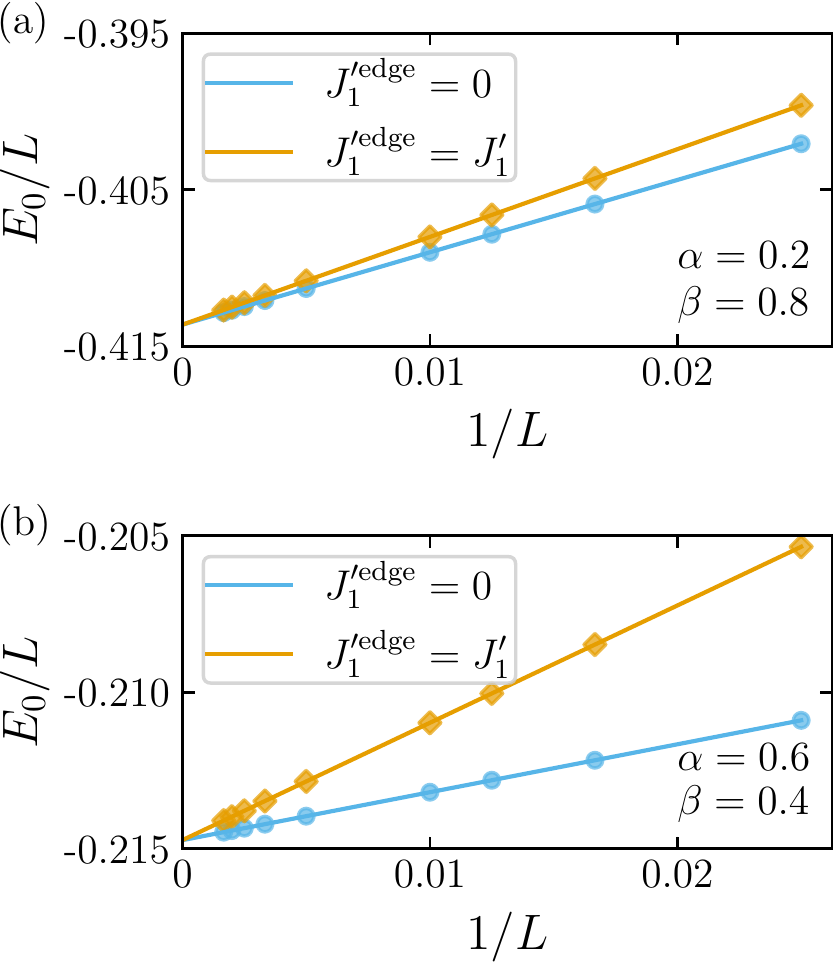}
	\caption{Finite-size scaling of the ground state energy per site $E_0/L$ for $J_{1}'^{\mathrm{edge}}$=0 (blue) and $J_{1}'^{\mathrm{edge}}=J'_1$ (orange) with (a) $\alpha$ = 0.8, $\beta$=0.2 and (b) $\alpha$ = 0.4, $\beta$=0.6.} 
	\label{fig:jedge}
\end{figure}

In this study, we applied special open boundary conditions where the $J_1'$ bonds at the chain edges are set to be zero, i.e.,  $J'^{\mathrm{edge}}_{1}=0$. This setting enables us to explicitly define the emergent spin-1 formations on the $J_1$ bonds, thereby offering us a mechanism to exert demonstrable control over the edge states of our Haldane-type VBS states. As a result, we are able to correctly calculate bulk physical quantities such as the spin gap as well as the SOP in the thermodynamic limit. In the absence of this setting, these quantities could be underestimated due to a potential confluence of two symmetry broken states. More details are discussed in Ref.~\onlinecite{Agrapidis2019}. To substantiate our hypothesis that the choice of boundary conditions does not affect the bulk ground state, we calculate the ground-state energies for the cases where $J'^{\mathrm{edge}}_{1}=0$ and $J'^{\mathrm{edge}}_{1}=J_1'$. Fig.~\ref{fig:jedge} shows a comparison of the energy per site ($E/L$) between these two scenarios for two sets of parameters ($\alpha$, $\beta$). For both parameter sets, it is observed that $E/L$ for $J'^{\mathrm{edge}}_{1}=0$ and $J'^{\mathrm{edge}}_{1}=J_1'$ seem to be extrapolated to the same value in the thermodynamic limit. The difference in $E/L$ at the thermodynamic limit between the two boundary conditions is found to be within the range of numerical zero, i.e., $\Delta E/L=10^{-8}-10^{-7}$. We thus conclude that the bulk ground state remains unaffected by the choice of $J'^{\mathrm{edge}}_{1}$.

\section{Finite-size scaling of the spin gap} \label{App:FSS}

\begin{figure}[tbh]
    \centering
    \includegraphics[width=0.8\columnwidth]{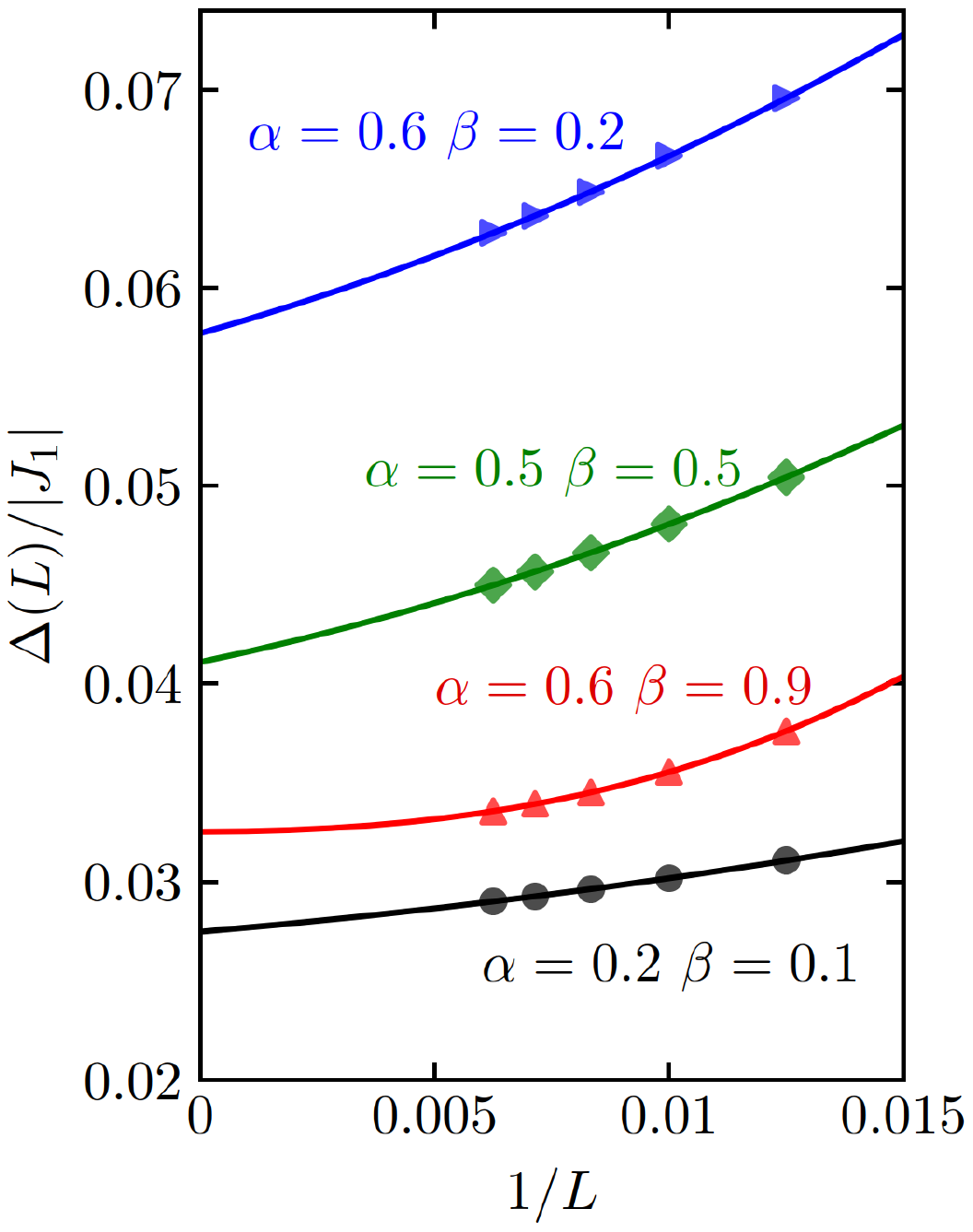}
    \caption{Examples of finite-size scaling analysis for the spin gap $\Delta(L)/|J_1|$, where open chains with lengths $L=80$, $100$, $120$, $140$, and $160$ are used.
  } 
    \label{fig:spingapscalex}
\end{figure}

In Fig.~\ref{fig:spingapscalex}, we show some examples of finite-size scaling analysis of the spin gap for various $(\alpha, \beta)$ values: $(0.6, 0.2)$ (blue line), $(0.5, 0.5)$ (green line), $(0.6, 0.9)$ (red line), and
$(0.2, 0.1)$ (black line). In most cases, the scaling can be reasonably performed by fitting the data with a quadratic function $\Delta(L)=a/L^2+b/L+\Delta$, where $a$ and $b$ are fitting parameters. In certain instances, particularly when the VBS state is quite robust, an alternative fitting function $\Delta(L)=a/L^3+b/L^2+\Delta$ is used. In such cases, the spin-spin correlation tends to decay rapidly with distance, leading to a fast convergence of the spin gap with respect to the system size. The parameter set $(\alpha, \beta)=(0.6, 0.9)$ is representative of such a case.

\section{Additional information about the entanglement spectrum}\label{appendix:ES}

\begin{figure}[t]
    \centering
    \includegraphics[width=0.8\columnwidth]{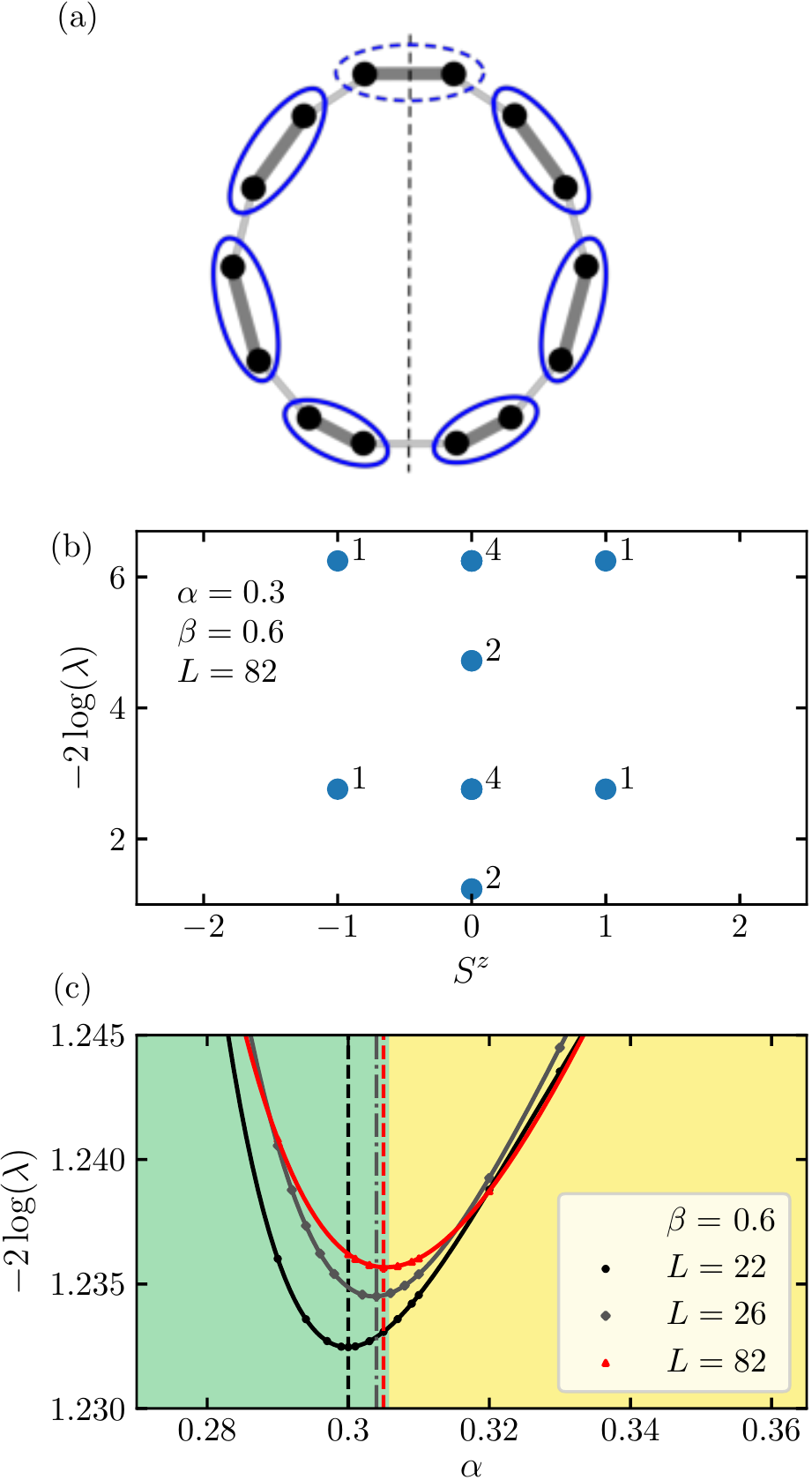}
    \caption{(a) Depiction of the partition of a periodic chain with 14 sites, where bold and thin lines represent $J_1$ and $J'_1$ interactions, respectively. The blue ellipses represent the emergent spin-1 sites, and the dashed-line ellipse denotes a fractionalized spin-1 site resulting from the partitioning. (b) Low-lying ES as a function of total $S^z$ of the subsystem for $(\alpha, \beta)=(0.3, 0.8)$ and $L=82$. The numbers next to each point indicate the degree of degeneracy. (c) Level of the lowest-lying 2-fold degenerate eigenstate in the ES as a function of $\alpha$ for $L=22$ (black), $L=26$ (gray), and $L=82$ (red) at $\beta=0.6$. Dashed vertical lines mark the position of the minimum for each $L$.} 
    \label{fig:ES_PBC_latt_Sz_res}
\end{figure}

We here provide additional information about our calculations for the ES. In our analysis for the ES the system size is chosen to be $L=4n+2=2(2n+1)$. Fig.~\ref{fig:ES_PBC_latt_Sz_res}(a) illustrates how the bipartite cutting acts on our periodic system when the ES is calculated. Indeed, the cut goes through two inequivalent bonds: one $J_1$ bond corresponding to the emerging spin-1 site, and one $J_1'$ bond. Thus, the cut creates two spin-1/2 edge states, one for each subsystem, leading to the 2-fold degeneracy of the lowest-lying eigenstate in the ES, as shown in Fig.~\ref{fig:ESalpha_beta_part}.

In Fig.~\ref{fig:ES_PBC_latt_Sz_res}(b) the low-lying ES for $(\alpha, \beta)=(0.3, 0.6)$ and $L=82$ is plotted as a function of total $S^z$ of the subsystem. we see that the 6-fold degeneracy in the second level is resolved to 4-fold degeneracy at $S^z=0$ and the remaining 2-fold degeneracy at $S^z=\pm1$. The states at  $S^z=\pm1$ may come from a possible spin-triplet components of each subsystem.

Moreover, as stated in the main text, the phase boundary between the two VBS states aligns with the minimum of the lowest-lying 2-fold degenerate eigenstate in the ES. To demonstrate this, we examine the behavior of the ES in the vicinity of the phase transition between the two VBS states. In Fig.~\ref{fig:ES_PBC_latt_Sz_res}(c), we plot the lowest-lying level of the ES as a function of $\alpha$ for system sizes $L=22$, $26$, and $82$ at a fixed $\beta=0.6$. It is apparent that the position of the minimum promptly approaches the phase boundary as the system size increases. This may be interpreted as the weight of the lowest-lying density-matrix eigenstate becoming more pronounced than those of the higher eigenstates at the phase boundary due to the robustness of the VBS state.

\begin{figure}[H]
	\centering
	\includegraphics[width=0.8\columnwidth]{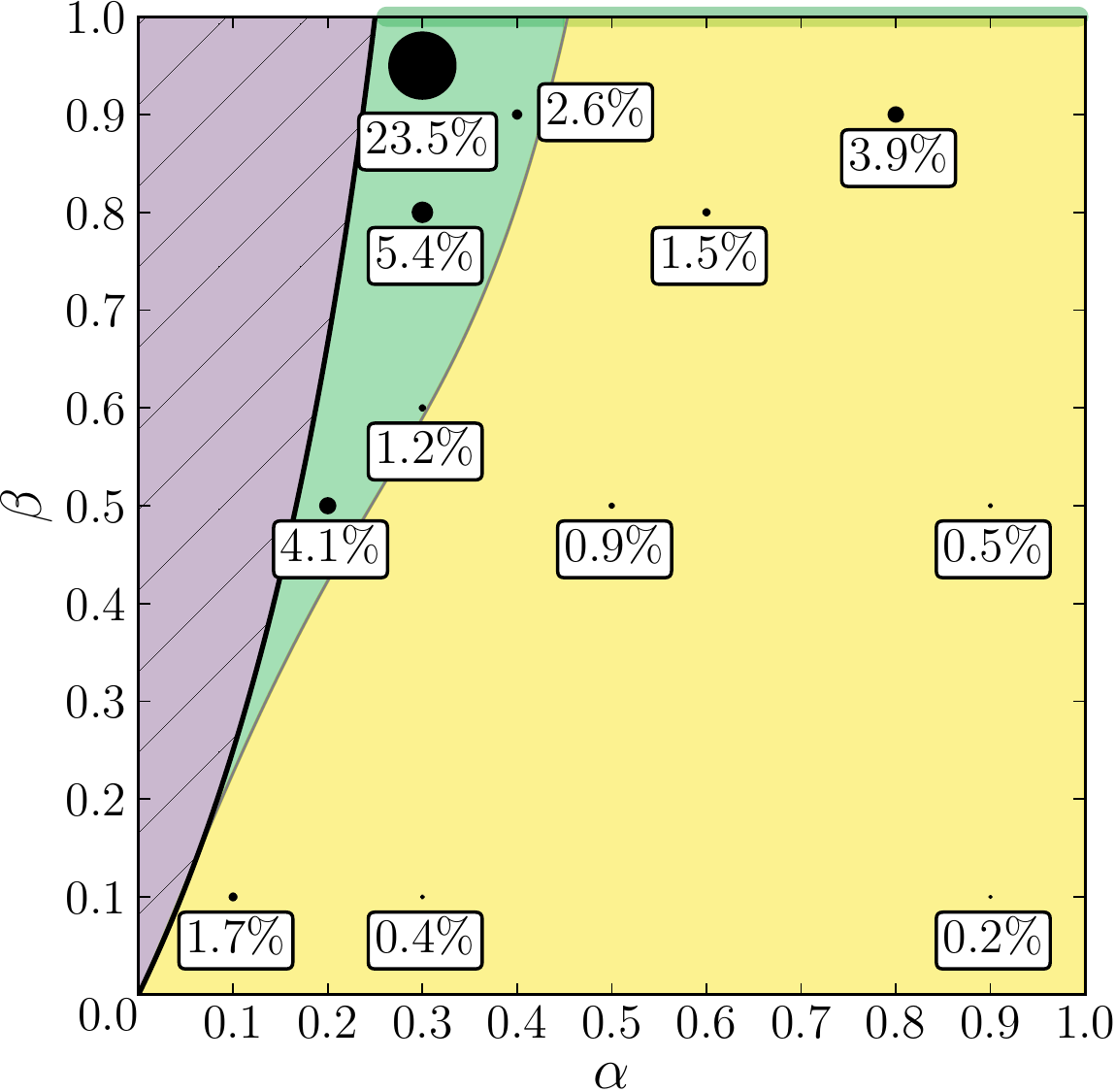}
	\caption{Relative increase of the spin gap with $J_3'$ in the $\alpha$-$\beta$ plane. The size of the circles is proportional to the relative increase of the gap at $J_3'=0.001|J_1|$. Results are superimposed to the phase diagram for clarity.} 
	\label{fig:J3percent}
\end{figure}

\section{Spin gap increase with $J'_3$}\label{App:GapJ3}

As discussed in section \ref{subsec:Gap_exp_dim_J3}, the spin gap exhibits a linear increase with the third-neighbor AFM interaction $J'_3$ within the mixed-VBS phase, and a 2/3 power-law relationship within the mixed-VBS phase. This section elaborates further on the disparate rates of increase observed in these two VBS phases. Fig.~\ref{fig:J3percent} presents the relative percentage increase of the spin gap when comparing $\Delta(J_3'=0)$ and $\Delta(J_3'=0.001|J_1|)$. The data reveals that the overall increase in the mixed-VBS region is of the order of $\sim1\%$. The growth of the spin gap is smaller if we increase the frustration parameter $\alpha$, while the opposite is true for the parameter $\beta$, with exceptions close to the critical line. Among the considered points, we see the most significant gap increase of $23.5\%$ for $(\alpha, \beta)=(0.3, 0.95)$. This increase can be ascribed to the third-neighbors valence bond geometry in the $\mathcal{D}_3$-VBS region. In the $\mathcal{D}_3$-VBS phase, the valence bonds are formed only on the $J_3'$ bonds, so that the increase of $J_3'$ directly enhances the spin gap. Whereas in the mixed-VBS phase, the spin gap is roughly estimated as an energy needed to break a second-neighbor valence bond. Thus, the increase of $J_3'$ just indirectly affects the spin gap.

\bibliography{references}

\begin{thebibliography}{47}%
\makeatletter
\providecommand \@ifxundefined [1]{%
 \@ifx{#1\undefined}
}%
\providecommand \@ifnum [1]{%
 \ifnum #1\expandafter \@firstoftwo
 \else \expandafter \@secondoftwo
 \fi
}%
\providecommand \@ifx [1]{%
 \ifx #1\expandafter \@firstoftwo
 \else \expandafter \@secondoftwo
 \fi
}%
\providecommand \natexlab [1]{#1}%
\providecommand \enquote  [1]{``#1''}%
\providecommand \bibnamefont  [1]{#1}%
\providecommand \bibfnamefont [1]{#1}%
\providecommand \citenamefont [1]{#1}%
\providecommand \href@noop [0]{\@secondoftwo}%
\providecommand \href [0]{\begingroup \@sanitize@url \@href}%
\providecommand \@href[1]{\@@startlink{#1}\@@href}%
\providecommand \@@href[1]{\endgroup#1\@@endlink}%
\providecommand \@sanitize@url [0]{\catcode `\\12\catcode `\$12\catcode
  `\&12\catcode `\#12\catcode `\^12\catcode `\_12\catcode `\%12\relax}%
\providecommand \@@startlink[1]{}%
\providecommand \@@endlink[0]{}%
\providecommand \url  [0]{\begingroup\@sanitize@url \@url }%
\providecommand \@url [1]{\endgroup\@href {#1}{\urlprefix }}%
\providecommand \urlprefix  [0]{URL }%
\providecommand \Eprint [0]{\href }%
\providecommand \doibase [0]{https://doi.org/}%
\providecommand \selectlanguage [0]{\@gobble}%
\providecommand \bibinfo  [0]{\@secondoftwo}%
\providecommand \bibfield  [0]{\@secondoftwo}%
\providecommand \translation [1]{[#1]}%
\providecommand \BibitemOpen [0]{}%
\providecommand \bibitemStop [0]{}%
\providecommand \bibitemNoStop [0]{.\EOS\space}%
\providecommand \EOS [0]{\spacefactor3000\relax}%
\providecommand \BibitemShut  [1]{\csname bibitem#1\endcsname}%
\let\auto@bib@innerbib\@empty
\bibitem [{\citenamefont {Balents}(2010)}]{Balents2010}%
  \BibitemOpen
  \bibfield  {author} {\bibinfo {author} {\bibfnamefont {L.}~\bibnamefont
  {Balents}},\ }\bibfield  {title} {\bibinfo {title} {Spin liquids in
  frustrated magnets},\ }\href@noop {} {\bibfield  {journal} {\bibinfo
  {journal} {Nature}\ }\textbf {\bibinfo {volume} {464}},\ \bibinfo {pages}
  {199} (\bibinfo {year} {2010})}\BibitemShut {NoStop}%
\bibitem [{\citenamefont {Moessner}\ and\ \citenamefont
  {Ramirez}(2006)}]{Moessner2006}%
  \BibitemOpen
  \bibfield  {author} {\bibinfo {author} {\bibfnamefont {R.}~\bibnamefont
  {Moessner}}\ and\ \bibinfo {author} {\bibfnamefont {A.~P.}\ \bibnamefont
  {Ramirez}},\ }\bibfield  {title} {\bibinfo {title} {{Geometrical
  frustration}},\ }\href {https://doi.org/10.1063/1.2186278} {\bibfield
  {journal} {\bibinfo  {journal} {Physics Today}\ }\textbf {\bibinfo {volume}
  {59}},\ \bibinfo {pages} {24} (\bibinfo {year} {2006})}\BibitemShut {NoStop}%
\bibitem [{\citenamefont {Kuzian}(2023)}]{Kuzian2023}%
  \BibitemOpen
  \bibfield  {author} {\bibinfo {author} {\bibfnamefont {R.}~\bibnamefont
  {Kuzian}},\ }\bibfield  {title} {\bibinfo {title} {Methods of modeling of
  strongly correlated electron systems},\ }\href
  {https://doi.org/10.3390/nano13020238} {\bibfield  {journal} {\bibinfo
  {journal} {Nanomaterials}\ }\textbf {\bibinfo {volume} {13}},\ \bibinfo
  {pages} {238} (\bibinfo {year} {2023})}\BibitemShut {NoStop}%
\bibitem [{\citenamefont {Majumdar}\ and\ \citenamefont
  {Ghosh}(1969)}]{Majumdar1969}%
  \BibitemOpen
  \bibfield  {author} {\bibinfo {author} {\bibfnamefont {C.~K.}\ \bibnamefont
  {Majumdar}}\ and\ \bibinfo {author} {\bibfnamefont {D.~K.}\ \bibnamefont
  {Ghosh}},\ }\bibfield  {title} {\bibinfo {title} {On next-nearest-neighbor
  interaction in linear chain. {I}},\ }\href@noop {} {\bibfield  {journal}
  {\bibinfo  {journal} {J. Math. Phys.}\ }\textbf {\bibinfo {volume} {10}},\
  \bibinfo {pages} {1388} (\bibinfo {year} {1969})}\BibitemShut {NoStop}%
\bibitem [{\citenamefont {Agrapidis}\ \emph {et~al.}(2019)\citenamefont
  {Agrapidis}, \citenamefont {Drechsler}, \citenamefont {van~den Brink},\ and\
  \citenamefont {Nishimoto}}]{Agrapidis2019}%
  \BibitemOpen
  \bibfield  {author} {\bibinfo {author} {\bibfnamefont {C.~E.}\ \bibnamefont
  {Agrapidis}}, \bibinfo {author} {\bibfnamefont {S.-L.}\ \bibnamefont
  {Drechsler}}, \bibinfo {author} {\bibfnamefont {J.}~\bibnamefont {van~den
  Brink}},\ and\ \bibinfo {author} {\bibfnamefont {S.}~\bibnamefont
  {Nishimoto}},\ }\bibfield  {title} {\bibinfo {title} {{Coexistence of
  valence-bond formation and topological order in the Frustrated Ferromagnetic
  $J_1$-$J_2$ Chain}},\ }\href {https://doi.org/10.21468/SciPostPhys.6.2.019}
  {\bibfield  {journal} {\bibinfo  {journal} {SciPost Phys.}\ }\textbf
  {\bibinfo {volume} {6}},\ \bibinfo {pages} {019} (\bibinfo {year}
  {2019})}\BibitemShut {NoStop}%
\bibitem [{\citenamefont {{H}aldane}(1983{\natexlab{a}})}]{Haldane1983a}%
  \BibitemOpen
  \bibfield  {author} {\bibinfo {author} {\bibfnamefont {F.}~\bibnamefont
  {{H}aldane}},\ }\bibfield  {title} {\bibinfo {title} {Continuum dynamics of
  the 1-{D} {H}eisenberg antiferromagnet: Identification with the ${O}(3)$
  nonlinear sigma model},\ }\href
  {https://doi.org/10.1016/0375-9601(83)90631-x} {\bibfield  {journal}
  {\bibinfo  {journal} {Phys. Lett. A}\ }\textbf {\bibinfo {volume} {93}},\
  \bibinfo {pages} {464–468} (\bibinfo {year}
  {1983}{\natexlab{a}})}\BibitemShut {NoStop}%
\bibitem [{\citenamefont {{H}aldane}(1983{\natexlab{b}})}]{Haldane1983b}%
  \BibitemOpen
  \bibfield  {author} {\bibinfo {author} {\bibfnamefont {F.~D.~M.}\
  \bibnamefont {{H}aldane}},\ }\bibfield  {title} {\bibinfo {title} {Nonlinear
  field theory of large-spin {H}eisenberg antiferromagnets: Semiclassically
  quantized solitons of the one-dimensional easy-axis {N}éel state},\ }\href
  {https://doi.org/10.1103/physrevlett.50.1153} {\bibfield  {journal} {\bibinfo
   {journal} {Phys. Rev. Lett.}\ }\textbf {\bibinfo {volume} {50}},\ \bibinfo
  {pages} {1153–1156} (\bibinfo {year} {1983}{\natexlab{b}})}\BibitemShut
  {NoStop}%
\bibitem [{\citenamefont {Botet}\ and\ \citenamefont
  {Jullien}(1983)}]{Botet1983}%
  \BibitemOpen
  \bibfield  {author} {\bibinfo {author} {\bibfnamefont {R.}~\bibnamefont
  {Botet}}\ and\ \bibinfo {author} {\bibfnamefont {R.}~\bibnamefont
  {Jullien}},\ }\bibfield  {title} {\bibinfo {title} {Ground-state properties
  of a spin-1 antiferromagnetic chain},\ }\href
  {https://doi.org/10.1103/physrevb.27.613} {\bibfield  {journal} {\bibinfo
  {journal} {Phys. Rev. B}\ }\textbf {\bibinfo {volume} {27}},\ \bibinfo
  {pages} {613–615} (\bibinfo {year} {1983})}\BibitemShut {NoStop}%
\bibitem [{\citenamefont {Renard}\ \emph {et~al.}(1988)\citenamefont {Renard},
  \citenamefont {Verdaguer}, \citenamefont {Regnault}, \citenamefont
  {Erkelens}, \citenamefont {Rossat-Mignod}, \citenamefont {Ribas},
  \citenamefont {Stirling},\ and\ \citenamefont {Vettier}}]{Renard1988}%
  \BibitemOpen
  \bibfield  {author} {\bibinfo {author} {\bibfnamefont {J.~P.}\ \bibnamefont
  {Renard}}, \bibinfo {author} {\bibfnamefont {M.}~\bibnamefont {Verdaguer}},
  \bibinfo {author} {\bibfnamefont {L.~P.}\ \bibnamefont {Regnault}}, \bibinfo
  {author} {\bibfnamefont {W.~A.~C.}\ \bibnamefont {Erkelens}}, \bibinfo
  {author} {\bibfnamefont {J.}~\bibnamefont {Rossat-Mignod}}, \bibinfo {author}
  {\bibfnamefont {J.}~\bibnamefont {Ribas}}, \bibinfo {author} {\bibfnamefont
  {W.~G.}\ \bibnamefont {Stirling}},\ and\ \bibinfo {author} {\bibfnamefont
  {C.}~\bibnamefont {Vettier}},\ }\bibfield  {title} {\bibinfo {title} {Quantum
  energy gap in two quasi-one-dimensional ${S}$=1 {H}eisenberg antiferromagnets
  (invited)},\ }\href {https://doi.org/10.1063/1.340736} {\bibfield  {journal}
  {\bibinfo  {journal} {J. of Appl. Phys.}\ }\textbf {\bibinfo {volume} {63}},\
  \bibinfo {pages} {3538–3542} (\bibinfo {year} {1988})}\BibitemShut
  {NoStop}%
\bibitem [{\citenamefont {Affleck}\ \emph {et~al.}(1987)\citenamefont
  {Affleck}, \citenamefont {Kennedy}, \citenamefont {Lieb},\ and\ \citenamefont
  {Tasaki}}]{Affleck1987}%
  \BibitemOpen
  \bibfield  {author} {\bibinfo {author} {\bibfnamefont {I.}~\bibnamefont
  {Affleck}}, \bibinfo {author} {\bibfnamefont {T.}~\bibnamefont {Kennedy}},
  \bibinfo {author} {\bibfnamefont {E.~H.}\ \bibnamefont {Lieb}},\ and\
  \bibinfo {author} {\bibfnamefont {H.}~\bibnamefont {Tasaki}},\ }\bibfield
  {title} {\bibinfo {title} {Rigorous results on valence-bond ground states in
  antiferromagnets},\ }\href {https://doi.org/10.1103/physrevlett.59.799}
  {\bibfield  {journal} {\bibinfo  {journal} {Phys. Rev. Lett.}\ }\textbf
  {\bibinfo {volume} {59}},\ \bibinfo {pages} {799–802} (\bibinfo {year}
  {1987})}\BibitemShut {NoStop}%
\bibitem [{\citenamefont {Affleck}\ \emph {et~al.}(1988)\citenamefont
  {Affleck}, \citenamefont {Kennedy}, \citenamefont {Lieb},\ and\ \citenamefont
  {Tasaki}}]{Affleck1988}%
  \BibitemOpen
  \bibfield  {author} {\bibinfo {author} {\bibfnamefont {I.}~\bibnamefont
  {Affleck}}, \bibinfo {author} {\bibfnamefont {T.}~\bibnamefont {Kennedy}},
  \bibinfo {author} {\bibfnamefont {E.~H.}\ \bibnamefont {Lieb}},\ and\
  \bibinfo {author} {\bibfnamefont {H.}~\bibnamefont {Tasaki}},\ }\bibfield
  {title} {\bibinfo {title} {Valence bond ground states in isotropic quantum
  antiferromagnets},\ }\href {https://doi.org/10.1007/bf01218021} {\bibfield
  {journal} {\bibinfo  {journal} {Comm. Math. Phys.}\ }\textbf {\bibinfo
  {volume} {115}},\ \bibinfo {pages} {477–528} (\bibinfo {year}
  {1988})}\BibitemShut {NoStop}%
\bibitem [{\citenamefont {Grafe}\ \emph {et~al.}(2017)\citenamefont {Grafe},
  \citenamefont {Nishimoto}, \citenamefont {Iakovleva}, \citenamefont
  {Vavilova}, \citenamefont {Spillecke}, \citenamefont {Alfonsov},
  \citenamefont {Sturza}, \citenamefont {Wurmehl}, \citenamefont {Nojiri},
  \citenamefont {Rosner}, \citenamefont {Richter}, \citenamefont
  {R\"{o}{\ss}ler}, \citenamefont {Drechsler}, \citenamefont {Kataev},\ and\
  \citenamefont {B\"{u}chner}}]{Grafe2017}%
  \BibitemOpen
  \bibfield  {author} {\bibinfo {author} {\bibfnamefont {H.-J.}\ \bibnamefont
  {Grafe}}, \bibinfo {author} {\bibfnamefont {S.}~\bibnamefont {Nishimoto}},
  \bibinfo {author} {\bibfnamefont {M.}~\bibnamefont {Iakovleva}}, \bibinfo
  {author} {\bibfnamefont {E.}~\bibnamefont {Vavilova}}, \bibinfo {author}
  {\bibfnamefont {L.}~\bibnamefont {Spillecke}}, \bibinfo {author}
  {\bibfnamefont {A.}~\bibnamefont {Alfonsov}}, \bibinfo {author}
  {\bibfnamefont {M.-I.}\ \bibnamefont {Sturza}}, \bibinfo {author}
  {\bibfnamefont {S.}~\bibnamefont {Wurmehl}}, \bibinfo {author} {\bibfnamefont
  {H.}~\bibnamefont {Nojiri}}, \bibinfo {author} {\bibfnamefont
  {H.}~\bibnamefont {Rosner}}, \bibinfo {author} {\bibfnamefont
  {J.}~\bibnamefont {Richter}}, \bibinfo {author} {\bibfnamefont {U.~K.}\
  \bibnamefont {R\"{o}{\ss}ler}}, \bibinfo {author} {\bibfnamefont {S.-L.}\
  \bibnamefont {Drechsler}}, \bibinfo {author} {\bibfnamefont {V.}~\bibnamefont
  {Kataev}},\ and\ \bibinfo {author} {\bibfnamefont {B.}~\bibnamefont
  {B\"{u}chner}},\ }\bibfield  {title} {\bibinfo {title} {Signatures of a
  magnetic field-induced unconventional nematic liquid in the frustrated and
  anisotropic spin-chain cuprate {LiCuSbO}4},\ }\href
  {https://doi.org/10.1038/s41598-017-06525-0} {\bibfield  {journal} {\bibinfo
  {journal} {Sci. Rep.}\ }\textbf {\bibinfo {volume} {7}},\ \bibinfo {pages}
  {6720} (\bibinfo {year} {2017})}\BibitemShut {NoStop}%
\bibitem [{\citenamefont {Hida}(1991)}]{Hida1991}%
  \BibitemOpen
  \bibfield  {author} {\bibinfo {author} {\bibfnamefont {K.}~\bibnamefont
  {Hida}},\ }\bibfield  {title} {\bibinfo {title} {{H}aldane gap in the
  spin$-1/2$ double chain {H}eisenberg antiferromagnet -numerical
  diagonalization and projector {M}onte {C}arlo study-},\ }\href
  {https://doi.org/10.1143/jpsj.60.1347} {\bibfield  {journal} {\bibinfo
  {journal} {J. Phys. Soc. Jpn.}\ }\textbf {\bibinfo {volume} {60}},\ \bibinfo
  {pages} {1347–1354} (\bibinfo {year} {1991})}\BibitemShut {NoStop}%
\bibitem [{\citenamefont {Watanabe}\ \emph {et~al.}(1993)\citenamefont
  {Watanabe}, \citenamefont {Nomura},\ and\ \citenamefont
  {Takada}}]{Watanabe1993}%
  \BibitemOpen
  \bibfield  {author} {\bibinfo {author} {\bibfnamefont {H.}~\bibnamefont
  {Watanabe}}, \bibinfo {author} {\bibfnamefont {K.}~\bibnamefont {Nomura}},\
  and\ \bibinfo {author} {\bibfnamefont {S.}~\bibnamefont {Takada}},\
  }\bibfield  {title} {\bibinfo {title} {${S}=1/2$ quantum {H}eisenberg ladder
  and ${S}=1$ {H}aldane phase},\ }\href {https://doi.org/10.1143/jpsj.62.2845}
  {\bibfield  {journal} {\bibinfo  {journal} {J. Phys. Soc. Jpn.}\ }\textbf
  {\bibinfo {volume} {62}},\ \bibinfo {pages} {2845–2860} (\bibinfo {year}
  {1993})}\BibitemShut {NoStop}%
\bibitem [{\citenamefont {White}\ and\ \citenamefont {Huse}(1993)}]{White1993}%
  \BibitemOpen
  \bibfield  {author} {\bibinfo {author} {\bibfnamefont {S.~R.}\ \bibnamefont
  {White}}\ and\ \bibinfo {author} {\bibfnamefont {D.~A.}\ \bibnamefont
  {Huse}},\ }\bibfield  {title} {\bibinfo {title} {Numerical
  renormalization-group study of low-lying eigenstates of the antiferromagnetic
  ${S}=1$ {H}eisenberg chain},\ }\href
  {https://doi.org/10.1103/physrevb.48.3844} {\bibfield  {journal} {\bibinfo
  {journal} {Phys. Rev. B}\ }\textbf {\bibinfo {volume} {48}},\ \bibinfo
  {pages} {3844} (\bibinfo {year} {1993})}\BibitemShut {NoStop}%
\bibitem [{\citenamefont {Schollw\"ock}(2005)}]{Schollwoeck2005}%
  \BibitemOpen
  \bibfield  {author} {\bibinfo {author} {\bibfnamefont {U.}~\bibnamefont
  {Schollw\"ock}},\ }\bibfield  {title} {\bibinfo {title} {The density-matrix
  renormalization group},\ }\href {https://doi.org/10.1103/RevModPhys.77.259}
  {\bibfield  {journal} {\bibinfo  {journal} {Rev. Mod. Phys.}\ }\textbf
  {\bibinfo {volume} {77}},\ \bibinfo {pages} {259} (\bibinfo {year}
  {2005})}\BibitemShut {NoStop}%
\bibitem [{\citenamefont {Agrapidis}\ \emph {et~al.}(2017)\citenamefont
  {Agrapidis}, \citenamefont {Drechsler}, \citenamefont {van~den Brink},\ and\
  \citenamefont {Nishimoto}}]{Agrapidis2017}%
  \BibitemOpen
  \bibfield  {author} {\bibinfo {author} {\bibfnamefont {C.~E.}\ \bibnamefont
  {Agrapidis}}, \bibinfo {author} {\bibfnamefont {S.-L.}\ \bibnamefont
  {Drechsler}}, \bibinfo {author} {\bibfnamefont {J.}~\bibnamefont {van~den
  Brink}},\ and\ \bibinfo {author} {\bibfnamefont {S.}~\bibnamefont
  {Nishimoto}},\ }\bibfield  {title} {\bibinfo {title} {Crossover from an
  incommensurate singlet spiral state with a vanishingly small spin gap to a
  valence-bond solid state in dimerized frustrated ferromagnetic spin chains},\
  }\href {https://doi.org/10.1103/PhysRevB.95.220404} {\bibfield  {journal}
  {\bibinfo  {journal} {Phys. Rev. B}\ }\textbf {\bibinfo {volume} {95}},\
  \bibinfo {pages} {220404} (\bibinfo {year} {2017})}\BibitemShut {NoStop}%
\bibitem [{\citenamefont {Oshikawa}(1992)}]{Oshikawa1992}%
  \BibitemOpen
  \bibfield  {author} {\bibinfo {author} {\bibfnamefont {M.}~\bibnamefont
  {Oshikawa}},\ }\bibfield  {title} {\bibinfo {title} {Hidden
  {$\mathbb{Z}_2\times\mathbb{Z}_2$} symmetry in quantum spin chains with
  arbitrary integer spin},\ }\href {https://doi.org/10.1088/0953-8984/4/36/019}
  {\bibfield  {journal} {\bibinfo  {journal} {J Phys. Condens. Matter}\
  }\textbf {\bibinfo {volume} {4}},\ \bibinfo {pages} {7469–7488} (\bibinfo
  {year} {1992})}\BibitemShut {NoStop}%
\bibitem [{\citenamefont {Kennedy}\ and\ \citenamefont
  {Tasaki}(1992{\natexlab{a}})}]{Kennedy1992a}%
  \BibitemOpen
  \bibfield  {author} {\bibinfo {author} {\bibfnamefont {T.}~\bibnamefont
  {Kennedy}}\ and\ \bibinfo {author} {\bibfnamefont {H.}~\bibnamefont
  {Tasaki}},\ }\bibfield  {title} {\bibinfo {title} {Hidden
  {$\mathbb{Z}_2\times\mathbb{Z}_2$} symmetry breaking in {H}aldane-gap
  antiferromagnets},\ }\href {https://doi.org/10.1103/physrevb.45.304}
  {\bibfield  {journal} {\bibinfo  {journal} {Phys. Rev. B}\ }\textbf {\bibinfo
  {volume} {45}},\ \bibinfo {pages} {304–307} (\bibinfo {year}
  {1992}{\natexlab{a}})}\BibitemShut {NoStop}%
\bibitem [{\citenamefont {Kennedy}\ and\ \citenamefont
  {Tasaki}(1992{\natexlab{b}})}]{Kennedy1992b}%
  \BibitemOpen
  \bibfield  {author} {\bibinfo {author} {\bibfnamefont {T.}~\bibnamefont
  {Kennedy}}\ and\ \bibinfo {author} {\bibfnamefont {H.}~\bibnamefont
  {Tasaki}},\ }\bibfield  {title} {\bibinfo {title} {Hidden symmetry breaking
  and the {H}aldane phase in ${S}=1$ quantum spin chains},\ }\href
  {https://doi.org/10.1007/bf02097239} {\bibfield  {journal} {\bibinfo
  {journal} {Comm. Math. Phys.}\ }\textbf {\bibinfo {volume} {147}},\ \bibinfo
  {pages} {431–484} (\bibinfo {year} {1992}{\natexlab{b}})}\BibitemShut
  {NoStop}%
\bibitem [{\citenamefont {Kennedy}(1990)}]{Kennedy1990}%
  \BibitemOpen
  \bibfield  {author} {\bibinfo {author} {\bibfnamefont {T.}~\bibnamefont
  {Kennedy}},\ }\bibfield  {title} {\bibinfo {title} {Exact diagonalisations of
  open spin-1 chains},\ }\href {https://doi.org/10.1088/0953-8984/2/26/010}
  {\bibfield  {journal} {\bibinfo  {journal} {J Phys. Condens. Matter}\
  }\textbf {\bibinfo {volume} {2}},\ \bibinfo {pages} {5737} (\bibinfo {year}
  {1990})}\BibitemShut {NoStop}%
\bibitem [{\citenamefont {Pollmann}\ \emph {et~al.}(2012)\citenamefont
  {Pollmann}, \citenamefont {Berg}, \citenamefont {Turner},\ and\ \citenamefont
  {Oshikawa}}]{Pollmann2012}%
  \BibitemOpen
  \bibfield  {author} {\bibinfo {author} {\bibfnamefont {F.}~\bibnamefont
  {Pollmann}}, \bibinfo {author} {\bibfnamefont {E.}~\bibnamefont {Berg}},
  \bibinfo {author} {\bibfnamefont {A.~M.}\ \bibnamefont {Turner}},\ and\
  \bibinfo {author} {\bibfnamefont {M.}~\bibnamefont {Oshikawa}},\ }\bibfield
  {title} {\bibinfo {title} {Symmetry protection of topological phases in
  one-dimensional quantum spin systems},\ }\href
  {https://doi.org/10.1103/physrevb.85.075125} {\bibfield  {journal} {\bibinfo
  {journal} {Phys. Rev. B}\ }\textbf {\bibinfo {volume} {85}},\ \bibinfo
  {pages} {075125} (\bibinfo {year} {2012})}\BibitemShut {NoStop}%
\bibitem [{\citenamefont {Gu}\ and\ \citenamefont {Wen}(2009)}]{Gu2009}%
  \BibitemOpen
  \bibfield  {author} {\bibinfo {author} {\bibfnamefont {Z.-C.}\ \bibnamefont
  {Gu}}\ and\ \bibinfo {author} {\bibfnamefont {X.-G.}\ \bibnamefont {Wen}},\
  }\bibfield  {title} {\bibinfo {title} {Tensor-entanglement-filtering
  renormalization approach and symmetry-protected topological order},\ }\href
  {https://doi.org/10.1103/physrevb.80.155131} {\bibfield  {journal} {\bibinfo
  {journal} {Phys. Rev. B}\ }\textbf {\bibinfo {volume} {80}},\ \bibinfo
  {pages} {155131} (\bibinfo {year} {2009})}\BibitemShut {NoStop}%
\bibitem [{\citenamefont {Tasaki}(2020)}]{Tasaki2020}%
  \BibitemOpen
  \bibfield  {author} {\bibinfo {author} {\bibfnamefont {H.}~\bibnamefont
  {Tasaki}},\ }\href {https://doi.org/10.1007/978-3-030-41265-4} {\emph
  {\bibinfo {title} {Physics and Mathematics of Quantum Many-Body Systems}}}\
  (\bibinfo  {publisher} {Springer International Publishing},\ \bibinfo
  {address} {Springer-Verlag, Berlin Heidelberg},\ \bibinfo {year}
  {2020})\BibitemShut {NoStop}%
\bibitem [{\citenamefont {den Nijs}\ and\ \citenamefont
  {Rommelse}(1989)}]{denNijs1989}%
  \BibitemOpen
  \bibfield  {author} {\bibinfo {author} {\bibfnamefont {M.}~\bibnamefont {den
  Nijs}}\ and\ \bibinfo {author} {\bibfnamefont {K.}~\bibnamefont {Rommelse}},\
  }\bibfield  {title} {\bibinfo {title} {Preroughening transitions in crystal
  surfaces and valence-bond phases in quantum spin chains},\ }\href
  {https://doi.org/10.1103/physrevb.40.4709} {\bibfield  {journal} {\bibinfo
  {journal} {Phys. Rev. B}\ }\textbf {\bibinfo {volume} {40}},\ \bibinfo
  {pages} {4709–4734} (\bibinfo {year} {1989})}\BibitemShut {NoStop}%
\bibitem [{\citenamefont {Kohmoto}\ and\ \citenamefont
  {Tasaki}(1992)}]{Kohmoto1992}%
  \BibitemOpen
  \bibfield  {author} {\bibinfo {author} {\bibfnamefont {M.}~\bibnamefont
  {Kohmoto}}\ and\ \bibinfo {author} {\bibfnamefont {H.}~\bibnamefont
  {Tasaki}},\ }\bibfield  {title} {\bibinfo {title} {Hidden ${S}=1/2$ quantum
  spin chain with bond alternation},\ }\href
  {https://doi.org/10.1103/physrevb.46.3486} {\bibfield  {journal} {\bibinfo
  {journal} {Phys. Rev. B}\ }\textbf {\bibinfo {volume} {46}},\ \bibinfo
  {pages} {3486–3495} (\bibinfo {year} {1992})}\BibitemShut {NoStop}%
\bibitem [{\citenamefont {Hida}(1992)}]{Hida1992}%
  \BibitemOpen
  \bibfield  {author} {\bibinfo {author} {\bibfnamefont {K.}~\bibnamefont
  {Hida}},\ }\bibfield  {title} {\bibinfo {title} {Crossover between the
  {H}aldane-gap phase and the dimer phase in the spin$-1/2$ alternating
  {H}eisenberg chain},\ }\href {https://doi.org/10.1103/physrevb.45.2207}
  {\bibfield  {journal} {\bibinfo  {journal} {Phys. Rev. B}\ }\textbf {\bibinfo
  {volume} {45}},\ \bibinfo {pages} {2207–2212} (\bibinfo {year}
  {1992})}\BibitemShut {NoStop}%
\bibitem [{\citenamefont {Pollmann}\ \emph {et~al.}(2010)\citenamefont
  {Pollmann}, \citenamefont {Turner}, \citenamefont {Berg},\ and\ \citenamefont
  {Oshikawa}}]{Pollmann2010}%
  \BibitemOpen
  \bibfield  {author} {\bibinfo {author} {\bibfnamefont {F.}~\bibnamefont
  {Pollmann}}, \bibinfo {author} {\bibfnamefont {A.~M.}\ \bibnamefont
  {Turner}}, \bibinfo {author} {\bibfnamefont {E.}~\bibnamefont {Berg}},\ and\
  \bibinfo {author} {\bibfnamefont {M.}~\bibnamefont {Oshikawa}},\ }\bibfield
  {title} {\bibinfo {title} {Entanglement spectrum of a topological phase in
  one dimension},\ }\href {https://doi.org/10.1103/physrevb.81.064439}
  {\bibfield  {journal} {\bibinfo  {journal} {Phys. Rev. B}\ }\textbf {\bibinfo
  {volume} {81}},\ \bibinfo {pages} {064439} (\bibinfo {year}
  {2010})}\BibitemShut {NoStop}%
\bibitem [{\citenamefont {Li}\ and\ \citenamefont {Li}(2008)}]{Li2008a}%
  \BibitemOpen
  \bibfield  {author} {\bibinfo {author} {\bibfnamefont {Y.-C.}\ \bibnamefont
  {Li}}\ and\ \bibinfo {author} {\bibfnamefont {S.-S.}\ \bibnamefont {Li}},\
  }\bibfield  {title} {\bibinfo {title} {Quantum phase transitions in the
  ${S}=1/2$ distorted diamond chain},\ }\href
  {https://doi.org/10.1103/physrevb.78.184412} {\bibfield  {journal} {\bibinfo
  {journal} {Phys. Rev. B}\ }\textbf {\bibinfo {volume} {78}},\ \bibinfo
  {pages} {184412} (\bibinfo {year} {2008})}\BibitemShut {NoStop}%
\bibitem [{\citenamefont {Fidkowski}(2010)}]{Fidkowski2010}%
  \BibitemOpen
  \bibfield  {author} {\bibinfo {author} {\bibfnamefont {L.}~\bibnamefont
  {Fidkowski}},\ }\bibfield  {title} {\bibinfo {title} {Entanglement spectrum
  of topological insulators and superconductors},\ }\href
  {https://doi.org/10.1103/physrevlett.104.130502} {\bibfield  {journal}
  {\bibinfo  {journal} {Phys. Rev. Lett.}\ }\textbf {\bibinfo {volume} {104}},\
  \bibinfo {pages} {130502} (\bibinfo {year} {2010})}\BibitemShut {NoStop}%
\bibitem [{\citenamefont {Li}\ and\ \citenamefont {Haldane}(2008)}]{Li2008b}%
  \BibitemOpen
  \bibfield  {author} {\bibinfo {author} {\bibfnamefont {H.}~\bibnamefont
  {Li}}\ and\ \bibinfo {author} {\bibfnamefont {F.~D.~M.}\ \bibnamefont
  {Haldane}},\ }\bibfield  {title} {\bibinfo {title} {Entanglement spectrum as
  a generalization of entanglement entropy: Identification of topological order
  in non-abelian fractional quantum hall effect states},\ }\href
  {https://doi.org/10.1103/physrevlett.101.010504} {\bibfield  {journal}
  {\bibinfo  {journal} {Phys. Rev. Lett.}\ }\textbf {\bibinfo {volume} {101}},\
  \bibinfo {pages} {010504} (\bibinfo {year} {2008})}\BibitemShut {NoStop}%
\bibitem [{\citenamefont {Vidal}\ \emph {et~al.}(2003)\citenamefont {Vidal},
  \citenamefont {Latorre}, \citenamefont {Rico},\ and\ \citenamefont
  {Kitaev}}]{Vidal2003}%
  \BibitemOpen
  \bibfield  {author} {\bibinfo {author} {\bibfnamefont {G.}~\bibnamefont
  {Vidal}}, \bibinfo {author} {\bibfnamefont {J.~I.}\ \bibnamefont {Latorre}},
  \bibinfo {author} {\bibfnamefont {E.}~\bibnamefont {Rico}},\ and\ \bibinfo
  {author} {\bibfnamefont {A.}~\bibnamefont {Kitaev}},\ }\bibfield  {title}
  {\bibinfo {title} {Entanglement in quantum critical phenomena},\ }\href
  {https://doi.org/10.1103/physrevlett.90.227902} {\bibfield  {journal}
  {\bibinfo  {journal} {Phys. Rev. Lett.}\ }\textbf {\bibinfo {volume} {90}},\
  \bibinfo {pages} {227902} (\bibinfo {year} {2003})}\BibitemShut {NoStop}%
\bibitem [{\citenamefont {Uhrig}\ \emph {et~al.}(1999)\citenamefont {Uhrig},
  \citenamefont {Sch\"{o}nfeld}, \citenamefont {Laukamp},\ and\ \citenamefont
  {Dagotto}}]{Uhrig1999}%
  \BibitemOpen
  \bibfield  {author} {\bibinfo {author} {\bibfnamefont {G.}~\bibnamefont
  {Uhrig}}, \bibinfo {author} {\bibfnamefont {F.}~\bibnamefont
  {Sch\"{o}nfeld}}, \bibinfo {author} {\bibfnamefont {M.}~\bibnamefont
  {Laukamp}},\ and\ \bibinfo {author} {\bibfnamefont {E.}~\bibnamefont
  {Dagotto}},\ }\bibfield  {title} {\bibinfo {title} {Unified quantum
  mechanical picture for confined spinons in dimerized and frustrated spin
  chains},\ }\href {https://doi.org/10.1007/s100510050589} {\bibfield
  {journal} {\bibinfo  {journal} {Eur. Phys. J. B}\ }\textbf {\bibinfo {volume}
  {7}},\ \bibinfo {pages} {67} (\bibinfo {year} {1999})}\BibitemShut {NoStop}%
\bibitem [{\citenamefont {Bosio{\v{c}}i{\'{c}}}\ \emph
  {et~al.}(2017)\citenamefont {Bosio{\v{c}}i{\'{c}}}, \citenamefont {Bert},
  \citenamefont {Dutton}, \citenamefont {Cava}, \citenamefont {Baker},
  \citenamefont {Po{\v{z}}ek},\ and\ \citenamefont {Mendels}}]{Bosioi2017}%
  \BibitemOpen
  \bibfield  {author} {\bibinfo {author} {\bibfnamefont {M.}~\bibnamefont
  {Bosio{\v{c}}i{\'{c}}}}, \bibinfo {author} {\bibfnamefont {F.}~\bibnamefont
  {Bert}}, \bibinfo {author} {\bibfnamefont {S.~E.}\ \bibnamefont {Dutton}},
  \bibinfo {author} {\bibfnamefont {R.~J.}\ \bibnamefont {Cava}}, \bibinfo
  {author} {\bibfnamefont {P.~J.}\ \bibnamefont {Baker}}, \bibinfo {author}
  {\bibfnamefont {M.}~\bibnamefont {Po{\v{z}}ek}},\ and\ \bibinfo {author}
  {\bibfnamefont {P.}~\bibnamefont {Mendels}},\ }\bibfield  {title} {\bibinfo
  {title} {Possible quadrupolar nematic phase in the frustrated spin chain
  {LiCuSbO}4: An nmr investigation},\ }\href
  {https://doi.org/10.1103/physrevb.96.224424} {\bibfield  {journal} {\bibinfo
  {journal} {Phys. Rev. B}\ }\textbf {\bibinfo {volume} {96}},\ \bibinfo
  {pages} {224424} (\bibinfo {year} {2017})}\BibitemShut {NoStop}%
\bibitem [{\citenamefont {Hikihara}\ \emph {et~al.}(2008)\citenamefont
  {Hikihara}, \citenamefont {Kecke}, \citenamefont {Momoi},\ and\ \citenamefont
  {Furusaki}}]{Hikihara2008}%
  \BibitemOpen
  \bibfield  {author} {\bibinfo {author} {\bibfnamefont {T.}~\bibnamefont
  {Hikihara}}, \bibinfo {author} {\bibfnamefont {L.}~\bibnamefont {Kecke}},
  \bibinfo {author} {\bibfnamefont {T.}~\bibnamefont {Momoi}},\ and\ \bibinfo
  {author} {\bibfnamefont {A.}~\bibnamefont {Furusaki}},\ }\bibfield  {title}
  {\bibinfo {title} {Vector chiral and multipolar orders in the spin-1/2
  frustrated ferromagnetic chain in magnetic field},\ }\href
  {https://doi.org/10.1103/physrevb.78.144404} {\bibfield  {journal} {\bibinfo
  {journal} {Phys. Rev. B}\ }\textbf {\bibinfo {volume} {78}},\ \bibinfo
  {pages} {144404} (\bibinfo {year} {2008})}\BibitemShut {NoStop}%
\bibitem [{\citenamefont {Sudan}\ \emph {et~al.}(2009)\citenamefont {Sudan},
  \citenamefont {L\"{u}scher},\ and\ \citenamefont {L\"{a}uchli}}]{Sudan2009}%
  \BibitemOpen
  \bibfield  {author} {\bibinfo {author} {\bibfnamefont {J.}~\bibnamefont
  {Sudan}}, \bibinfo {author} {\bibfnamefont {A.}~\bibnamefont {L\"{u}scher}},\
  and\ \bibinfo {author} {\bibfnamefont {A.~M.}\ \bibnamefont {L\"{a}uchli}},\
  }\bibfield  {title} {\bibinfo {title} {Emergent multipolar spin correlations
  in a fluctuating spiral: The frustrated ferromagnetic spin-1/2 {H}eisenberg
  chain in a magnetic field},\ }\href
  {https://doi.org/10.1103/physrevb.80.140402} {\bibfield  {journal} {\bibinfo
  {journal} {Phys. Rev. B}\ }\textbf {\bibinfo {volume} {80}},\ \bibinfo
  {pages} {140402} (\bibinfo {year} {2009})}\BibitemShut {NoStop}%
\bibitem [{\citenamefont {Mourigal}\ \emph {et~al.}(2012)\citenamefont
  {Mourigal}, \citenamefont {Enderle}, \citenamefont {F{\aa}k}, \citenamefont
  {Kremer}, \citenamefont {Law}, \citenamefont {Schneidewind}, \citenamefont
  {Hiess},\ and\ \citenamefont {Prokofiev}}]{Mourigal2012}%
  \BibitemOpen
  \bibfield  {author} {\bibinfo {author} {\bibfnamefont {M.}~\bibnamefont
  {Mourigal}}, \bibinfo {author} {\bibfnamefont {M.}~\bibnamefont {Enderle}},
  \bibinfo {author} {\bibfnamefont {B.}~\bibnamefont {F{\aa}k}}, \bibinfo
  {author} {\bibfnamefont {R.~K.}\ \bibnamefont {Kremer}}, \bibinfo {author}
  {\bibfnamefont {J.~M.}\ \bibnamefont {Law}}, \bibinfo {author} {\bibfnamefont
  {A.}~\bibnamefont {Schneidewind}}, \bibinfo {author} {\bibfnamefont
  {A.}~\bibnamefont {Hiess}},\ and\ \bibinfo {author} {\bibfnamefont
  {A.}~\bibnamefont {Prokofiev}},\ }\bibfield  {title} {\bibinfo {title}
  {Evidence of a bond-nematic phase in {L}i{C}u{VO}4},\ }\href
  {https://doi.org/10.1103/physrevlett.109.027203} {\bibfield  {journal}
  {\bibinfo  {journal} {Phys. Rev. Lett.}\ }\textbf {\bibinfo {volume} {109}},\
  \bibinfo {pages} {027203} (\bibinfo {year} {2012})}\BibitemShut {NoStop}%
\bibitem [{\citenamefont {Vishwanath}\ \emph {et~al.}(2004)\citenamefont
  {Vishwanath}, \citenamefont {Balents},\ and\ \citenamefont
  {Senthil}}]{Vishwanath2004}%
  \BibitemOpen
  \bibfield  {author} {\bibinfo {author} {\bibfnamefont {A.}~\bibnamefont
  {Vishwanath}}, \bibinfo {author} {\bibfnamefont {L.}~\bibnamefont
  {Balents}},\ and\ \bibinfo {author} {\bibfnamefont {T.}~\bibnamefont
  {Senthil}},\ }\bibfield  {title} {\bibinfo {title} {Quantum criticality and
  deconfinement in phase transitions between valence bond solids},\ }\href
  {https://doi.org/10.1103/physrevb.69.224416} {\bibfield  {journal} {\bibinfo
  {journal} {Phys. Rev. B}\ }\textbf {\bibinfo {volume} {69}},\ \bibinfo
  {pages} {224416} (\bibinfo {year} {2004})}\BibitemShut {NoStop}%
\bibitem [{\citenamefont {Slagle}\ and\ \citenamefont {Xu}(2014)}]{Slagle2014}%
  \BibitemOpen
  \bibfield  {author} {\bibinfo {author} {\bibfnamefont {K.}~\bibnamefont
  {Slagle}}\ and\ \bibinfo {author} {\bibfnamefont {C.}~\bibnamefont {Xu}},\
  }\bibfield  {title} {\bibinfo {title} {Quantum phase transition between the
  {Z}$_2$ spin liquid and valence bond crystals on a triangular lattice},\
  }\href {https://doi.org/10.1103/physrevb.89.104418} {\bibfield  {journal}
  {\bibinfo  {journal} {Phys. Rev. B}\ }\textbf {\bibinfo {volume} {89}},\
  \bibinfo {pages} {104418} (\bibinfo {year} {2014})}\BibitemShut {NoStop}%
\bibitem [{\citenamefont {Liao}\ \emph {et~al.}(2022)\citenamefont {Liao},
  \citenamefont {Xu}, \citenamefont {Meng},\ and\ \citenamefont
  {Qi}}]{DaLiao2022}%
  \BibitemOpen
  \bibfield  {author} {\bibinfo {author} {\bibfnamefont {Y.~D.}\ \bibnamefont
  {Liao}}, \bibinfo {author} {\bibfnamefont {X.~Y.}\ \bibnamefont {Xu}},
  \bibinfo {author} {\bibfnamefont {Z.~Y.}\ \bibnamefont {Meng}},\ and\
  \bibinfo {author} {\bibfnamefont {Y.}~\bibnamefont {Qi}},\ }\bibfield
  {title} {\bibinfo {title} {Dirac fermions with plaquette interactions. iii.
  {SU(N)} phase diagram with {G}ross-{N}eveu criticality and first-order phase
  transition},\ }\href {https://doi.org/10.1103/physrevb.106.155159} {\bibfield
   {journal} {\bibinfo  {journal} {Phys. Rev. B}\ }\textbf {\bibinfo {volume}
  {106}},\ \bibinfo {pages} {155159} (\bibinfo {year} {2022})}\BibitemShut
  {NoStop}%
\bibitem [{\citenamefont {Tonegawa}\ \emph {et~al.}(1998)\citenamefont
  {Tonegawa}, \citenamefont {Hikihara}, \citenamefont {Kaburagi}, \citenamefont
  {Nishino}, \citenamefont {Miyashita},\ and\ \citenamefont
  {Mikeska}}]{Tonegawa1998}%
  \BibitemOpen
  \bibfield  {author} {\bibinfo {author} {\bibfnamefont {T.}~\bibnamefont
  {Tonegawa}}, \bibinfo {author} {\bibfnamefont {T.}~\bibnamefont {Hikihara}},
  \bibinfo {author} {\bibfnamefont {M.}~\bibnamefont {Kaburagi}}, \bibinfo
  {author} {\bibfnamefont {T.}~\bibnamefont {Nishino}}, \bibinfo {author}
  {\bibfnamefont {S.}~\bibnamefont {Miyashita}},\ and\ \bibinfo {author}
  {\bibfnamefont {H.-J.}\ \bibnamefont {Mikeska}},\ }\bibfield  {title}
  {\bibinfo {title} {Ground-state and thermodynamic properties of the quantum
  mixed spin-1/2-1/2-1-1 chain},\ }\href {https://doi.org/10.1143/jpsj.67.1000}
  {\bibfield  {journal} {\bibinfo  {journal} {J. Phys. Soc. Jpn}\ }\textbf
  {\bibinfo {volume} {67}},\ \bibinfo {pages} {1000} (\bibinfo {year}
  {1998})}\BibitemShut {NoStop}%
\bibitem [{\citenamefont {Li}\ \emph {et~al.}(2006)\citenamefont {Li},
  \citenamefont {Xu}, \citenamefont {Dai},\ and\ \citenamefont {Xu}}]{Li2006}%
  \BibitemOpen
  \bibfield  {author} {\bibinfo {author} {\bibfnamefont {S.-B.}\ \bibnamefont
  {Li}}, \bibinfo {author} {\bibfnamefont {Z.-X.}\ \bibnamefont {Xu}}, \bibinfo
  {author} {\bibfnamefont {J.-H.}\ \bibnamefont {Dai}},\ and\ \bibinfo {author}
  {\bibfnamefont {J.-B.}\ \bibnamefont {Xu}},\ }\bibfield  {title} {\bibinfo
  {title} {Entanglement and quantum phase transitions in quantum mixed spin
  chains},\ }\href {https://doi.org/10.1103/physrevb.73.184411} {\bibfield
  {journal} {\bibinfo  {journal} {Phys. Rev. B}\ }\textbf {\bibinfo {volume}
  {73}},\ \bibinfo {pages} {184411} (\bibinfo {year} {2006})}\BibitemShut
  {NoStop}%
\bibitem [{\citenamefont {Ver{\'{\i}}ssimo}\ \emph {et~al.}(2023)\citenamefont
  {Ver{\'{\i}}ssimo}, \citenamefont {Pereira}, \citenamefont {Stre{\v{c}}ka},\
  and\ \citenamefont {Lyra}}]{Verssimo2023}%
  \BibitemOpen
  \bibfield  {author} {\bibinfo {author} {\bibfnamefont {L.~M.}\ \bibnamefont
  {Ver{\'{\i}}ssimo}}, \bibinfo {author} {\bibfnamefont {M.~S.}\ \bibnamefont
  {Pereira}}, \bibinfo {author} {\bibfnamefont {J.}~\bibnamefont
  {Stre{\v{c}}ka}},\ and\ \bibinfo {author} {\bibfnamefont {M.~L.}\
  \bibnamefont {Lyra}},\ }\bibfield  {title} {\bibinfo {title} {Topological
  quantum phase transition in a mixed-spin {H}eisenberg tetramer chain with
  alternating spin-1/2 and spin-5/2 dimers},\ }\href
  {https://doi.org/10.1016/j.jmmm.2023.170595} {\bibfield  {journal} {\bibinfo
  {journal} {J. Magn. Magn. Mater.}\ }\textbf {\bibinfo {volume} {571}},\
  \bibinfo {pages} {170595} (\bibinfo {year} {2023})}\BibitemShut {NoStop}%
\bibitem [{\citenamefont {Kitazawa}\ and\ \citenamefont
  {Nomura}(1997)}]{Kitazawa1997}%
  \BibitemOpen
  \bibfield  {author} {\bibinfo {author} {\bibfnamefont {A.}~\bibnamefont
  {Kitazawa}}\ and\ \bibinfo {author} {\bibfnamefont {K.}~\bibnamefont
  {Nomura}},\ }\bibfield  {title} {\bibinfo {title} {Phase transitions of
  ${S}=3/2$ and ${S}=2$ {XXZ} spin chains with bond alternation},\ }\href
  {https://doi.org/10.1143/jpsj.66.3379} {\bibfield  {journal} {\bibinfo
  {journal} {J. Phys. Soc. Jpn}\ }\textbf {\bibinfo {volume} {66}},\ \bibinfo
  {pages} {3379} (\bibinfo {year} {1997})}\BibitemShut {NoStop}%
\bibitem [{\citenamefont {Miyakoshi}\ \emph {et~al.}(2016)\citenamefont
  {Miyakoshi}, \citenamefont {Nishimoto},\ and\ \citenamefont
  {Ohta}}]{Miyakoshi2016}%
  \BibitemOpen
  \bibfield  {author} {\bibinfo {author} {\bibfnamefont {S.}~\bibnamefont
  {Miyakoshi}}, \bibinfo {author} {\bibfnamefont {S.}~\bibnamefont
  {Nishimoto}},\ and\ \bibinfo {author} {\bibfnamefont {Y.}~\bibnamefont
  {Ohta}},\ }\bibfield  {title} {\bibinfo {title} {Entanglement properties of
  the {H}aldane phases: A finite system-size approach},\ }\href
  {https://doi.org/10.1103/physrevb.94.235155} {\bibfield  {journal} {\bibinfo
  {journal} {Phys. Rev. B}\ }\textbf {\bibinfo {volume} {94}},\ \bibinfo
  {pages} {235155} (\bibinfo {year} {2016})}\BibitemShut {NoStop}%
\bibitem [{\citenamefont {Hase}\ \emph {et~al.}(2004)\citenamefont {Hase},
  \citenamefont {Kuroe}, \citenamefont {Ozawa}, \citenamefont {Suzuki},
  \citenamefont {Kitazawa}, \citenamefont {Kido},\ and\ \citenamefont
  {Sekine}}]{Hase2004}%
  \BibitemOpen
  \bibfield  {author} {\bibinfo {author} {\bibfnamefont {M.}~\bibnamefont
  {Hase}}, \bibinfo {author} {\bibfnamefont {H.}~\bibnamefont {Kuroe}},
  \bibinfo {author} {\bibfnamefont {K.}~\bibnamefont {Ozawa}}, \bibinfo
  {author} {\bibfnamefont {O.}~\bibnamefont {Suzuki}}, \bibinfo {author}
  {\bibfnamefont {H.}~\bibnamefont {Kitazawa}}, \bibinfo {author}
  {\bibfnamefont {G.}~\bibnamefont {Kido}},\ and\ \bibinfo {author}
  {\bibfnamefont {T.}~\bibnamefont {Sekine}},\ }\bibfield  {title} {\bibinfo
  {title} {Magnetic properties of
  ${\mathrm{rb}}_{2}{\mathrm{cu}}_{2}{\mathrm{mo}}_{3}{\mathrm{o}}_{12}$
  including a one-dimensional spin-$1/2$ heisenberg system with ferromagnetic
  first-nearest-neighbor and antiferromagnetic second-nearest-neighbor exchange
  interactions},\ }\href {https://doi.org/10.1103/PhysRevB.70.104426}
  {\bibfield  {journal} {\bibinfo  {journal} {Phys. Rev. B}\ }\textbf {\bibinfo
  {volume} {70}},\ \bibinfo {pages} {104426} (\bibinfo {year}
  {2004})}\BibitemShut {NoStop}%
\bibitem [{\citenamefont {Wardyn}\ \emph {et~al.}(2023)\citenamefont {Wardyn},
  \citenamefont {Agrapidis},\ and\ \citenamefont {Nishimoto}}]{zenodo}%
  \BibitemOpen
  \bibfield  {author} {\bibinfo {author} {\bibfnamefont {J.}~\bibnamefont
  {Wardyn}}, \bibinfo {author} {\bibfnamefont {C.~E.}\ \bibnamefont
  {Agrapidis}},\ and\ \bibinfo {author} {\bibfnamefont {S.}~\bibnamefont
  {Nishimoto}},\ }\bibfield  {title} {\bibinfo {title} {Existence of two
  distinct valence bond solid states in the dimerized frustrated ferromagnetic
  ${J}_1$-${J}'_1$-${J}_2$ chain},\ }\bibfield  {journal} {\bibinfo  {journal}
  {zenodo.7997521}\ }\href {https://doi.org/10.5281/ZENODO.7997521}
  {10.5281/ZENODO.7997521} (\bibinfo {year} {2023})\BibitemShut {NoStop}%
\end{thebibliography}%

\end{document}